\documentclass[preprint,aps,superscript,address,showkeys,showpacs,pre]{revtex4}
\usepackage{amsmath,amssymb}
\usepackage{epsfig}
\usepackage{graphics}
\usepackage{color}
\usepackage{graphicx, lineno}

\begin{document}
%
%
\title{Synchronization of two non-identical Chialvo neurons}

\author{Javier Used}
\email[]{javier.used@urjc.es}
\affiliation{Nonlinear Dynamics, Chaos and Complex Systems Group, Departamento de
F\'{i}sica, Universidad Rey Juan Carlos, Tulip\'{a}n s/n, 28933 M\'{o}stoles, Madrid, Spain}

\author{Jes\'{u}s M. Seoane}
\affiliation{Nonlinear Dynamics, Chaos and Complex Systems Group, Departamento de
F\'{i}sica, Universidad Rey Juan Carlos, Tulip\'{a}n s/n, 28933 M\'{o}stoles, Madrid, Spain}

\author{Irina Bashkirtseva}
\affiliation{Institute of Natural Sciences and Mathematics, Ural Federal University, 620000, Lenina, 51, Ekaterinburg, Russia}

\author{Lev Ryashko}
\affiliation{Institute of Natural Sciences and Mathematics, Ural Federal University, 620000, Lenina, 51, Ekaterinburg, Russia}

\author{Miguel A.F. Sanju\'{a}n}
\affiliation{Nonlinear Dynamics, Chaos and Complex Systems Group, Departamento de
F\'{i}sica, Universidad Rey Juan Carlos, Tulip\'{a}n s/n, 28933 M\'{o}stoles, Madrid, Spain}

\date{\today}

\begin{abstract}

We investigate the synchronization between two neurons using the stochastic version of the map-based Chialvo model. To simulate non-identical neurons, a mismatch is introduced in one of the main parameters of the model. Subsequently, the synchronization of the neurons is studied as a function of this mismatch, the noise introduced in the stochastic model, and the coupling strength between the neurons. We propose the simplest neural network for study, as its analysis is more straightforward and does not compromise generality. Within this network, two non-identical neurons are electrically coupled. In order to understand whether specific behaviors affect the global behavior of the system, we consider different cases related to the behavior of the neurons (chaotic or periodic). Furthermore, we study how variations in model parameters affect the firing frequency in each case. Additionally, we consider that the two neurons have both excitatory and inhibitory couplings. Consequently, we identify critical values of noise and mismatch for achieving satisfactory synchronization between the neurons in each case. Finally, we propose that the results have general applicability across various neuron models.

\end{abstract}
\maketitle
\newpage
\section{Introduction} \label{introduction}
The topic of modeling neural dynamics is a relevant research field in Nonlinear Dynamics and Chaos and it has applications and implications for biological systems. One of the pioneering works on this topic was carried out by Hodgkin and Huxley that describes the relationship between action potentials and the ion currents \cite{Hodgkin1952,Hodgkin1952_2}.
The analysis of this system can be very complex, and therefore simpler models were developed for that purpose. Some of the most relevant models of smaller dimensions included the FitzHugh-Nagumo \cite{FitzHugh1961}, the Morris-Lecar \cite{morris:lecar} and the Hindmarsch-Rose \cite{Hindmarsh1982, Shilnikov2008} models, among others. All these models are continuous models and therefore can be modeled using ordinary differential equations and are well known in the literature.

On the other hand, the modeling of this kind of system using maps has been quite relevant and an important step since they can be mathematically treated and analyzed more easily. In this sense, the discrete model of Rulkov \cite{Rulkov01} is one of the most characteristic and simpler discrete models that mimics the dynamics of a neuron, but many different models of neurons in the form of maps of various dimensions are also well known in the literature \cite{Ibarz2011,Gir2013,KT96}. All these maps are quite useful for the study of neural networks, synchronization phenomena, among others, in the context of excitable systems. \textcolor{red}{In this work, we apply the Chialvo model \cite{Chialvo95} as the prototype to simulate the behavior of the neurons.} Although until now this model has not received much attention, in recent years several authors have focused their efforts to try to understand the complexity that this model is able to reproduce \cite{WANG2018,Yang2020}. Recently, it has  been studied in the context of noise-induced synchronization \cite{bruss:2023} by using an stochastic version of the model.

The effect of noise in excitable systems has been considerably studied for white and coloured noise  \cite{zambrano:2010}. In this work, the authors also considered the presence of intrinsic noise as well as the existence of a parameter mismatch between the neurons finding a resonant-like behavior when the neurons are identical and, therefore, the parameter mismatch is zero. This last work was conducted for the situation of a network of $2$ neurons modeled by the FitzHugh-Nagumo system.

Although the study of networks of neurons based on modified Chialvo neuron model has received some attention in recent \cite{SRIRAM2023, Muni_2022,Vivekanandhan_2023}, there has not been work in the literature focused on the connection of neurons based in the stochastic Chialvo model. For this reason in this work, we focus on the effects of parameter mismatch between neurons in a two-neuron network of Chialvo maps in presence of white Gaussian noise by using numerical methods. We also consider that the neurons are coupled by both excitatory coupling and inhibitory coupling. The type of coupling between the neurons plays a crucial role because \textcolor{red}{when a neuron fires and has excitatory coupling with the other it activates the other neurons and propagates electrical signals throughout the brain}. The inhibitory coupling does the opposite, so it makes the rest of the neurons less likely to fire messages of their own. The goal is to analyze how they affect both the synchronization rate and the inter-spike interval ($ISI$). Since the Chialvo model is a prototype model that exhibits all the main characteristic of a neural model, we conjecture that similar phenomena occur in other models both continuous and discrete.

The organization of this paper is as follows. In Section~\ref{modeldescription1N}, we describe the neuron Chialvo model. In Section~\ref{2neuronsection}, we analyze the synchronization phenomenon on the coupled $2$-neuron network by using excitatory and inhibitory couplings.  The main conclusions and a discussion of our results are provided in Section~\ref{conclusions}.

\section{The map-based Chialvo model}
 \label{modeldescription1N}

We start describing the stochastic version of the map-based Chialvo model \cite{Chialvo95} that is given by
\begin{equation}\label{model1}
\begin{array}{l}
  x_{t+1}=x_t^2 \exp(y_t-x_t)+I+\varepsilon\xi_t,\medskip\\
  y_{t+1}=a y_t-b x_t +c.
\end{array}
\end{equation}

In this model, the variable $x$ is related to an instantaneous membrane potential of the neuron while the variable $y$ stands for the recovery current. The parameter $I$ models the action of the ion current injected into the neuron, while the parameter $a$ is related to the time of recovery ($a < 1$), and the activation-dependence of the recovery process is defined by the parameter $b$  ($b < 1$) with the offset value $c$. In the stochastic model, a random disturbance of intensity $\varepsilon\xi_t$ is added to the parameter $I$ of the acting ion current. The parameter $\varepsilon$ represents the maximum strength of the noise, while $\xi_t$ corresponds to an uncorrelated white Gaussian noise with parameters $\langle\xi_t\rangle=0,\;\langle\xi^2_t\rangle=1$.

The original individual neuron model, where noise is not present, is capable of exhibiting rich dynamics and could adequately reproduce some of the basic features behind the firing dynamics of the neurons \cite{Chialvo95, Hoppensteadt86}. For example, by modifying the parameter $b$, one could obtain different dynamical behaviors: fixed points, periodic, quasi-periodic or chaotic dynamics. The value of the parameter $b$ is also related to the inter-spike interval, and depending of its value, the model can reproduce chaotic or fixed inter-spike intervals. We can observe an increase of the inter-spike intervals as the parameter $b$ increases, resulting in a decrease in the firing frequency of the neuron.

For relatively small values of $b$, the system possesses a stable equilibrium. However, as $b$ increases, this equilibrium loses its stability due to a Neimark-Sacker bifurcation. Consequently, attractors in the form of closed invariant curves emerge, which later are destroyed giving rise to strange attractors. Finally, for high values of $b$, limit cycles appear.

\begin{figure}[htb!]
\centering
\includegraphics[width=1\textwidth]{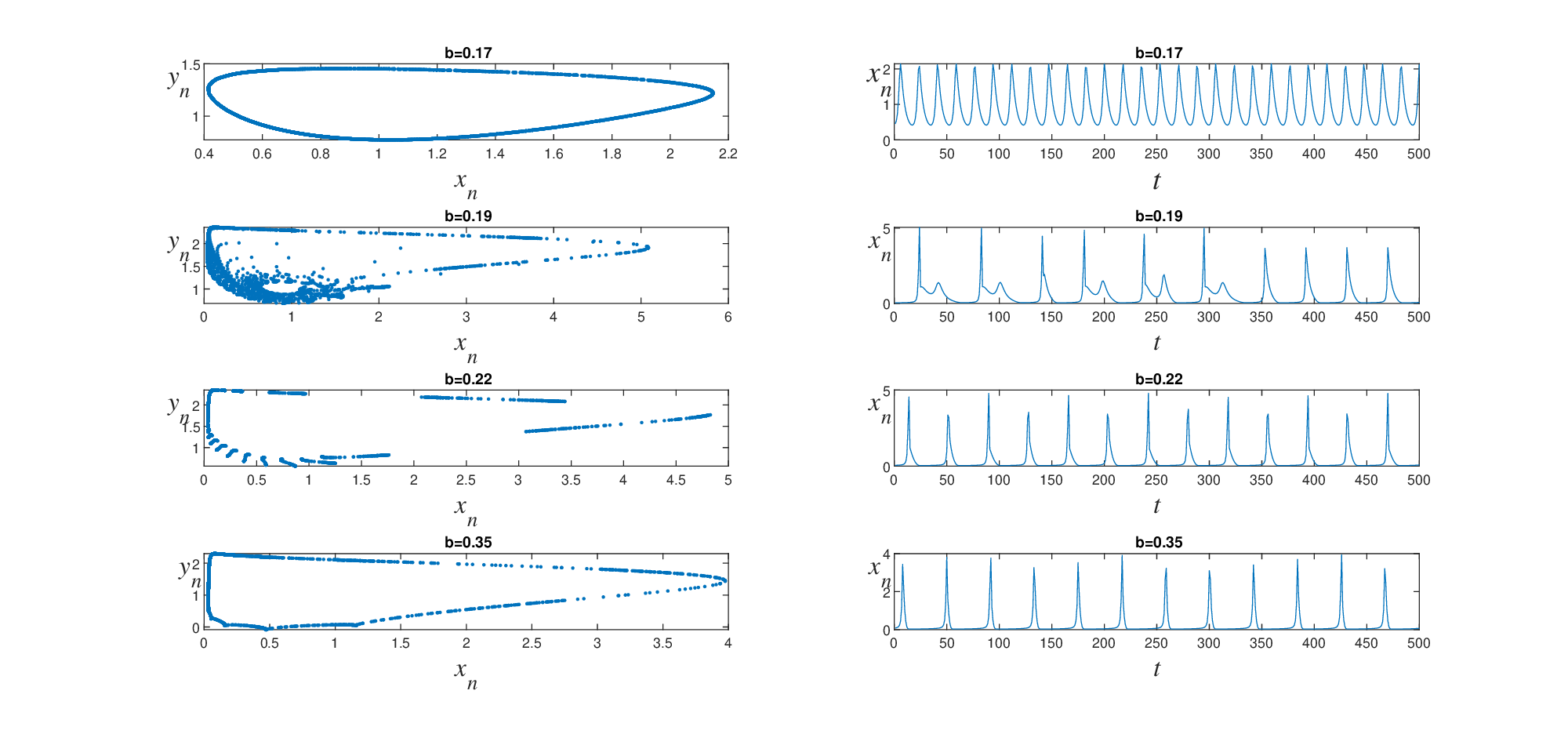}
\caption{These pictures show the attractors of the chaotic system (left panel) and the time series of the variable $x_n$ (right panel) for different values of the parameter $b$ according to the inset. The rest of the model parameters are fixed as $a=0.89,\;I=0.03,\;c=0.28$, and $\varepsilon=0.001$. As the value of the parameter $b$ changes, the model reproduces different behaviors. These changes in the attractors and in the time series of the variable $x_n$ can be appreciated as parameter $b$ changes.}
\label{space_phase_time_series_1N_Db}
\end{figure}

\begin{figure}[htb!]
\centering
\includegraphics[width=1\textwidth]{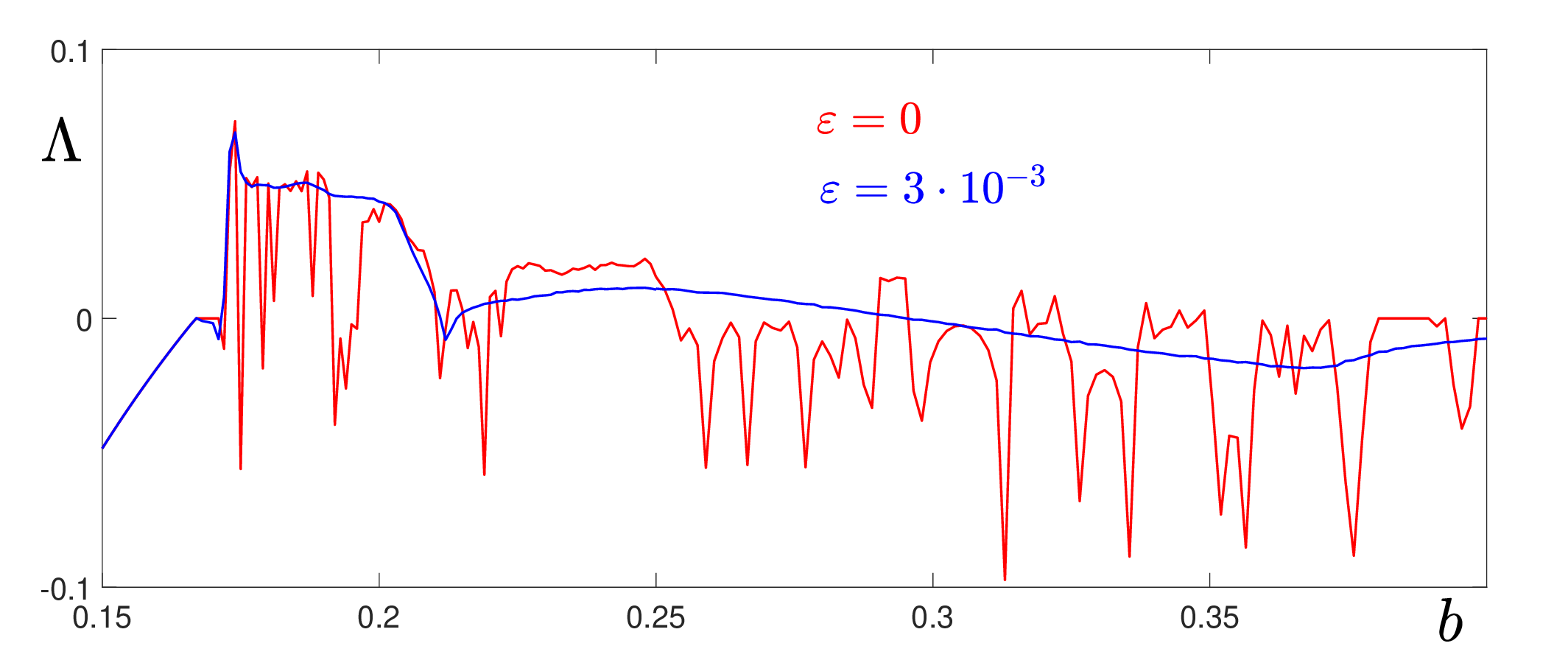}
\caption{Largest Lyapunov exponent for the system (Eq.~\ref{model1}) versus the parameter $b$: red for $\varepsilon=0$ and blue for $\varepsilon=0.003$. The rest of the model parameters are fixed to $a=0.89,\;I=0.03,$ and $\;c=0.28$. We can observe the effects of noise since as the curve $\Lambda(b)$ becomes smoother, reducing the order-chaos transitions.}
\label{Largest_Lyapunov}
\end{figure}

The diversity of these regimes can also be analyzed in Fig.~\ref{Largest_Lyapunov} where the largest Lyapunov exponent $\Lambda(b)$ is plotted in red for $\varepsilon=0$ and in blue for $\varepsilon=0.003$. Here, we have fixed $a=0.89,\;I=0.03,\;c=0.28$ and we have considered $b$ as the bifurcation parameter. We can also observe in this figure the chaotic behaviors where $\Lambda > 0$, plotted as a function of the parameter $b$. As can be seen, a variation of $b$ implies crucial changes in the system dynamics (Eq.~\ref{model1}) with order-chaos transitions. The effect of introducing noise as a random perturbation smoothen the graph $\Lambda(b)$. This smoothing reduces the order-chaos transitions, that is, the noise stabilizes the neuron behavior.

In Figs.~\ref{space_phase_time_series_1N_Db} and ~\ref{Largest_Lyapunov}, attractors with their time series and their corresponding maximum Lyapunov exponent can be analyzed for four values of the parameter $b$. For $b=0.17$, the attractor of the deterministic system is the stable closed invariant curve with a limit cycle ($\Lambda=0$).  For $b=0.19$, we have the chaotic attractor with complex geometry ($\Lambda=0.052>0$). Besides, for $b=0.22$, we also have the chaotic attractor ($\Lambda=0.0079>0$), but with another distribution of states, and finally $b=0.35$, the stable discrete $42$-cycle ($\Lambda=-0.018<0$) is an attractor of the system (Eq.~\ref{model1}).

\begin{figure}[htb!]
\centering
a) \includegraphics[width=0.7\textwidth]{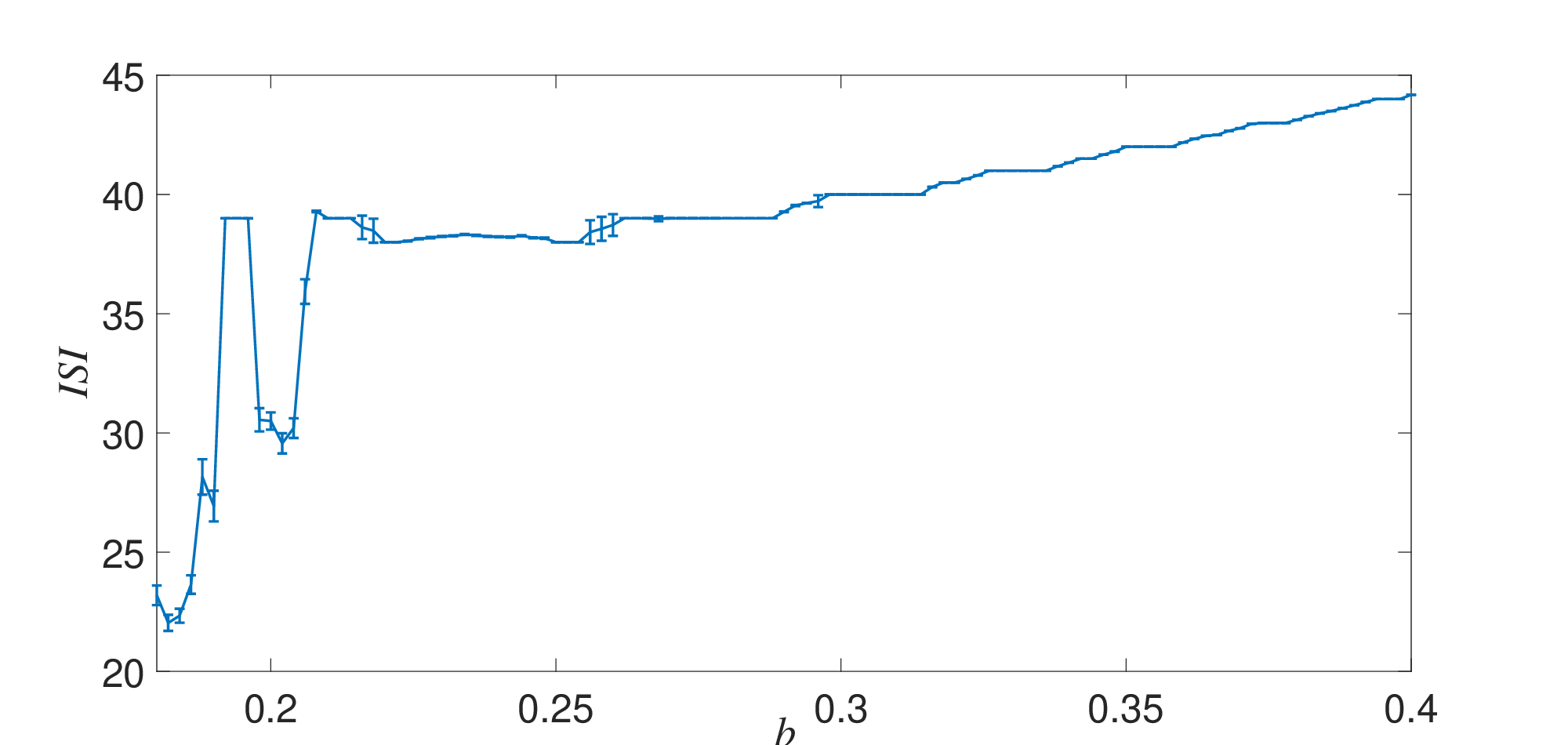}

b) \includegraphics[width=0.7\textwidth]{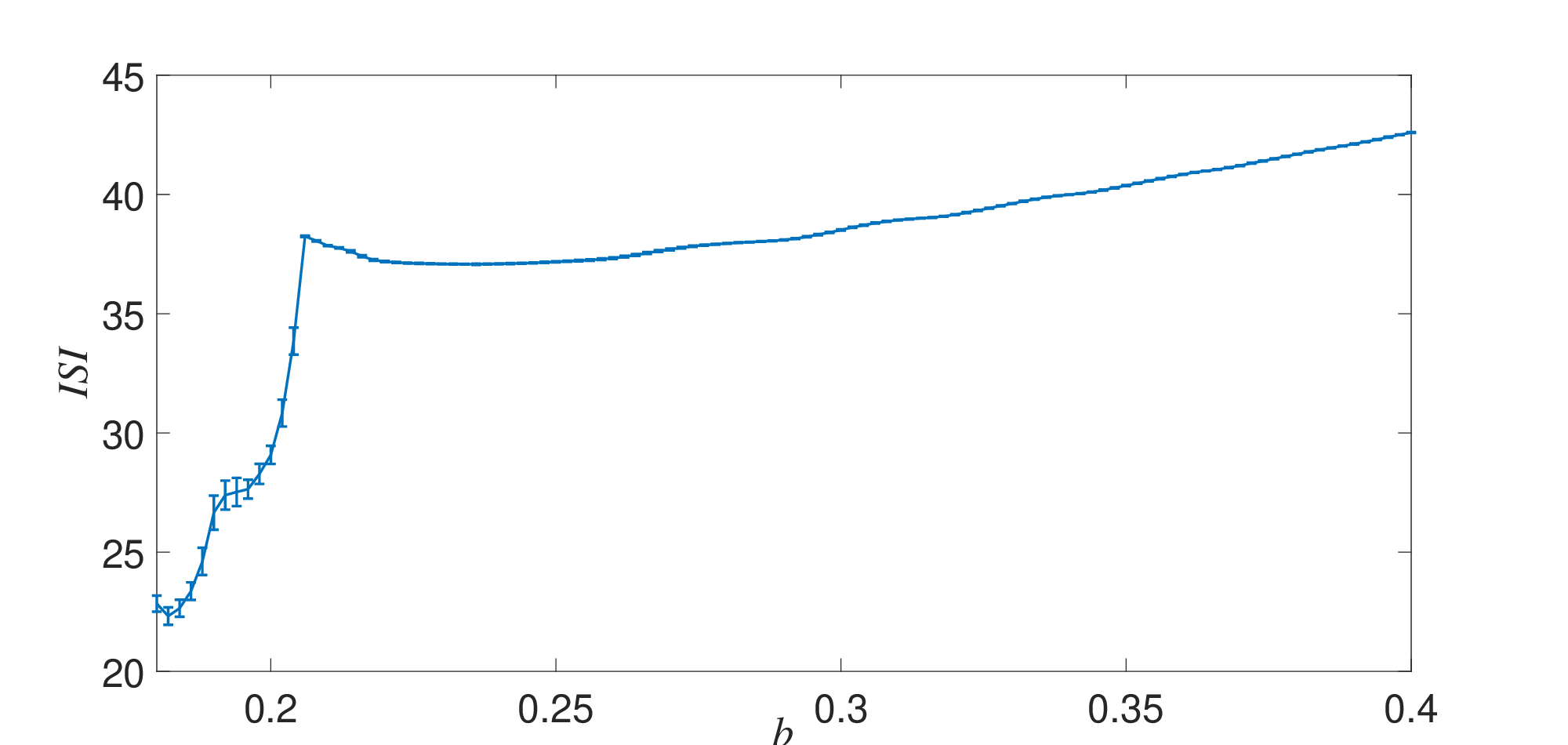}
 \caption{Inter-spike interval ($ISI$) as a function of the parameter $b$. The $ISI$ and its standard deviation are plotted as a function of the parameter $b$. The rest of the parameters are constant values $a=0.89$, $c=0.28$ and $I=0.03$. In the upper figure, (a), no noise has been added, while in the lower figure (b), the parameter $\varepsilon$ has a value of $1.5\cdot 10^{-3}$. Each point on the figure represents the mean value of $50$ simulations, each one starting from random initial conditions.}
\label{fig_figura_ISI_b_a_89_prueba}
\end{figure}

Now, we plot Fig.~\ref{fig_figura_ISI_b_a_89_prueba} in order to analyze the effect of noise on the global behavior of the system. In Fig.~\ref{fig_figura_ISI_b_a_89_prueba}, the inter-spike interval ($ISI$), which is the time between two consecutive maxima, and its standard deviation, are plotted also as a function of the parameter $b$. A brief analysis of this figure allows us to observe the effect of noise on the system's behavior and the slight differences between both models. In this way, it can be observed how the presence of noise in the system stabilizes the firing frequency of the neuron. Additionally, it can be also observed that the noise reduces slightly the mean value of the $ISI$ and also reduces its standard deviation.

From a neurophysiological point of view, the analysis of the $ISI$ is particularly interesting because some neurobiologists conjecture that the firing frequency of the neuron can be related to neural coding \cite{Fujii96, MacLeod96}. Some differences between Figs.~\ref{fig_figura_ISI_b_a_89_prueba}(a) and (b) can be appreciated. In Fig.~\ref{fig_figura_ISI_b_a_89_prueba}(a), corresponding to the noiseless model, the values of the $ISI$ reach higher values than in the stochastic model (Fig.~\ref{fig_figura_ISI_b_a_89_prueba}(b)). This trend is observable for lower values of $b$ and becomes more significant for higher values. For instance, the $ISI$ takes a value of $\sim 75$  when $b=0.6$ in the noiseless model, while it takes a value of $\sim 55$ in the stochastic model. Another difference between these cases is observed around the region $0.19<b<0.21$. In the noiseless model, a large variation of the $ISI$ is observable depending on the value of the parameter $b$. For example, when $b=0.19$ the value of $ISI$ is $\sim 27$. Suddenly, the value of $ISI$ increases to $39$ for the parameter values $0.192\le b \le 0.196$. Later, it abruptly decreases for $0.198 \le b \le 0.206$ and takes a value of $\sim 30$. From that point, the value of $ISI$ experiences a slight decrease before gradually increasing as $b$ increases. This tendency is observable in both models.

A very remarkable feature in both models, is the stability of the $ISI$. In Fig.~\ref{fig_figura_ISI_b_a_89_prueba}, we can observe that the standard deviation of the $ISI$ for $b>0.206$ is very close to zero for both models. In fact, in the noiseless model there are many values for which the standard deviation is zero. In the stochastic model, the values of the standard deviation are also very close to zero. This means that despite the differences in the amplitude that the model reproduces, it is a very stable dynamical system in the frequency domain.

\textcolor{red}{Based on the information extracted from the previous results}, we will focus on the last three cases ($b=0.19,\;b=0.22$ and $b=0.35$). The first two cases correspond to regions where the system exhibits chaotic behavior and the $ISI$ of the neuron changes significantly with small variations in the parameter $b$. In the last case, the model shows a more stable response and the firing frequency remains almost constant for similar values of the parameter $b$. In all these cases, the response of a  two-neuron system will be studied as a function of the noise intensity, the mismatch in the parameter $b$ of both neurons and the type of coupling, aiming to understand the role that these parameters play in the stability of the system.

\section{The two-neuron network}\label{2neuronsection}
Having describe the single-one neuron model,  we now turn to the study of a two-neuron system coupled by electrical means. We examine the effects of noise on the behavior of the single-neuron model. While this model represents the simplest unit for study, we believe it serves as a valuable starting point for analyzing whether noise can also play a significant role in the global synchronization of neurons.  As mentioned in the Introduction (Sec. \ref{introduction}), to introduce inhomogeneity into the model, we propose a mismatch between the parameters $b$ of neuron $1$ and neuron $2$. Hence, we can consider that for neuron $2$ its parameter $b$ can be written as $b_2=b_1+\Delta b$, where $\Delta b \ll b_{1}, b_{2}$, so that the equations of the model are:
\begin{equation}\label{model2}
\begin{array}{l}
  x_{1,t+1}=x_{1,t}^2 \exp(y_{1,t}-x_{1,t})+I \pm k(x_{2,t}-x_{1,t})+\varepsilon\xi_{1,t}\medskip\\
  y_{1,t+1}=a y_{1,t}-b_1 x_{1,t} +c\medskip\\
  x_{2,t+1}=x_{2,t}^2 \exp(y_{2,t}-x_{2,t})+I \pm k(x_{1,t}-x_{2,t})+\varepsilon\xi_{2,t}\medskip\\
  y_{2,t+1}=a y_{2,t}-(b_1+\Delta b)  x_{2,t} +c,
\end{array}
\end{equation}
where the sign of the coupling coefficient $k$ represents the two types of coupling that can exist between neurons: excitatory ($+$) and inhibitory ($-$)\cite{TISA:2006}. The random value of the noise in the model is different for each neuron in every step and it will be reproduced by means of the random values of $\xi_{1,t}$ and $\xi_{2,t}$, respectively.\\

Our goal is to provide a quantitative measure of the synchronization between the variable $x$ for both neurons ($x_1$ and $x_2$). For that purpose, we use the order parameter $R$ ~\cite{Ojalvo93} whose mathematical expression is given by:
	\begin{equation}\notag
		R(x)=\frac{\left<\overline{x}^2\right>-\left<\overline{x}\right>^2}{\overline{\left<x^2\right>-\left<x\right>^2}},
	\end{equation}\label{Requation}
where $\overline{x}$ is the mean average over all neurons, and $\left<x\right>$ is the mean in time, i.e., the mean of the variable $x$ of each neuron.  Accordingly, $R(x)\in\left[0,1\right]$, where $R(x)=0$ means total de-synchronization and $R(x)=1$ means total synchronization of the time series $x$.

\begin{figure}[ht!]
\centering
a)  \includegraphics[width=0.9\textwidth]{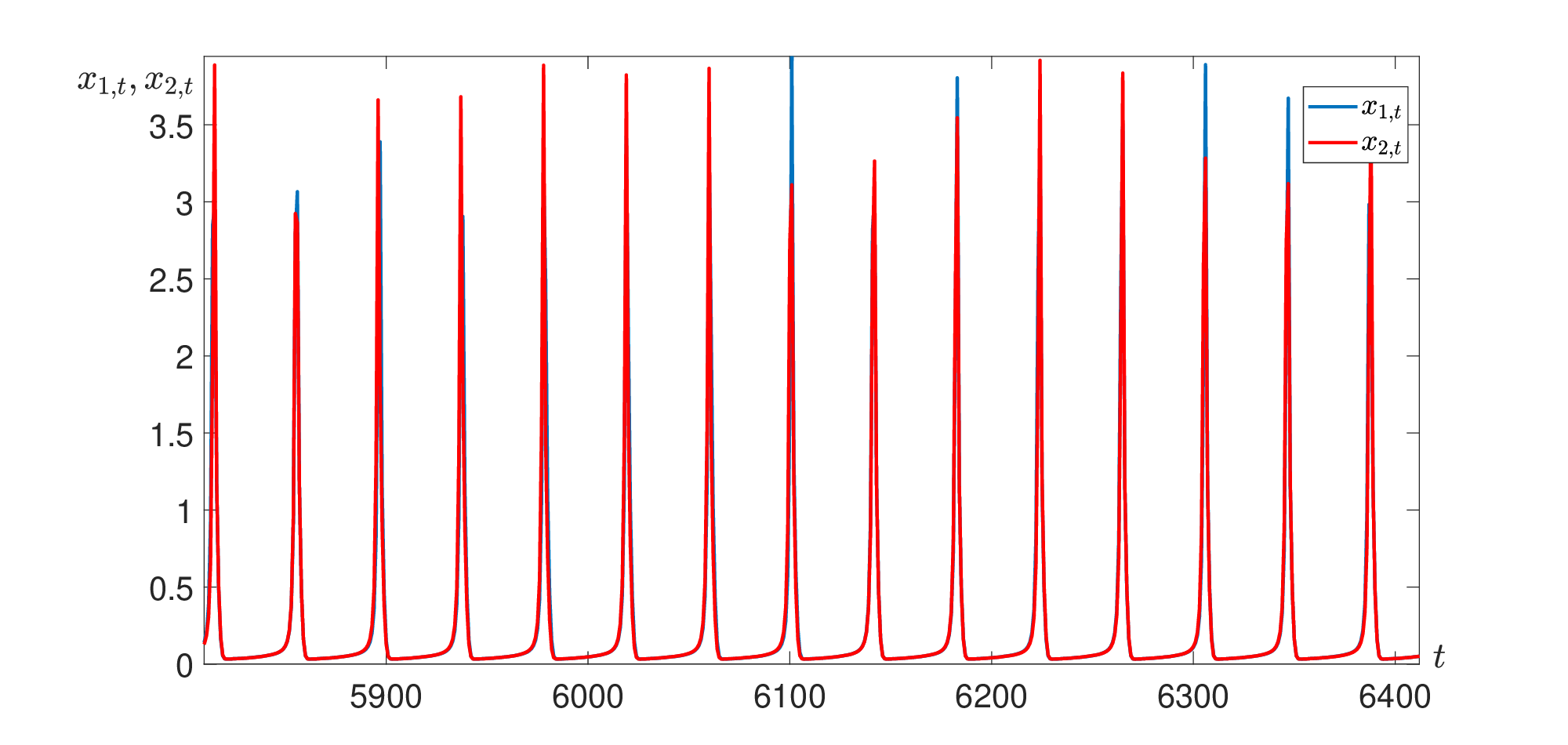}

b)  \includegraphics[width=0.9\textwidth]{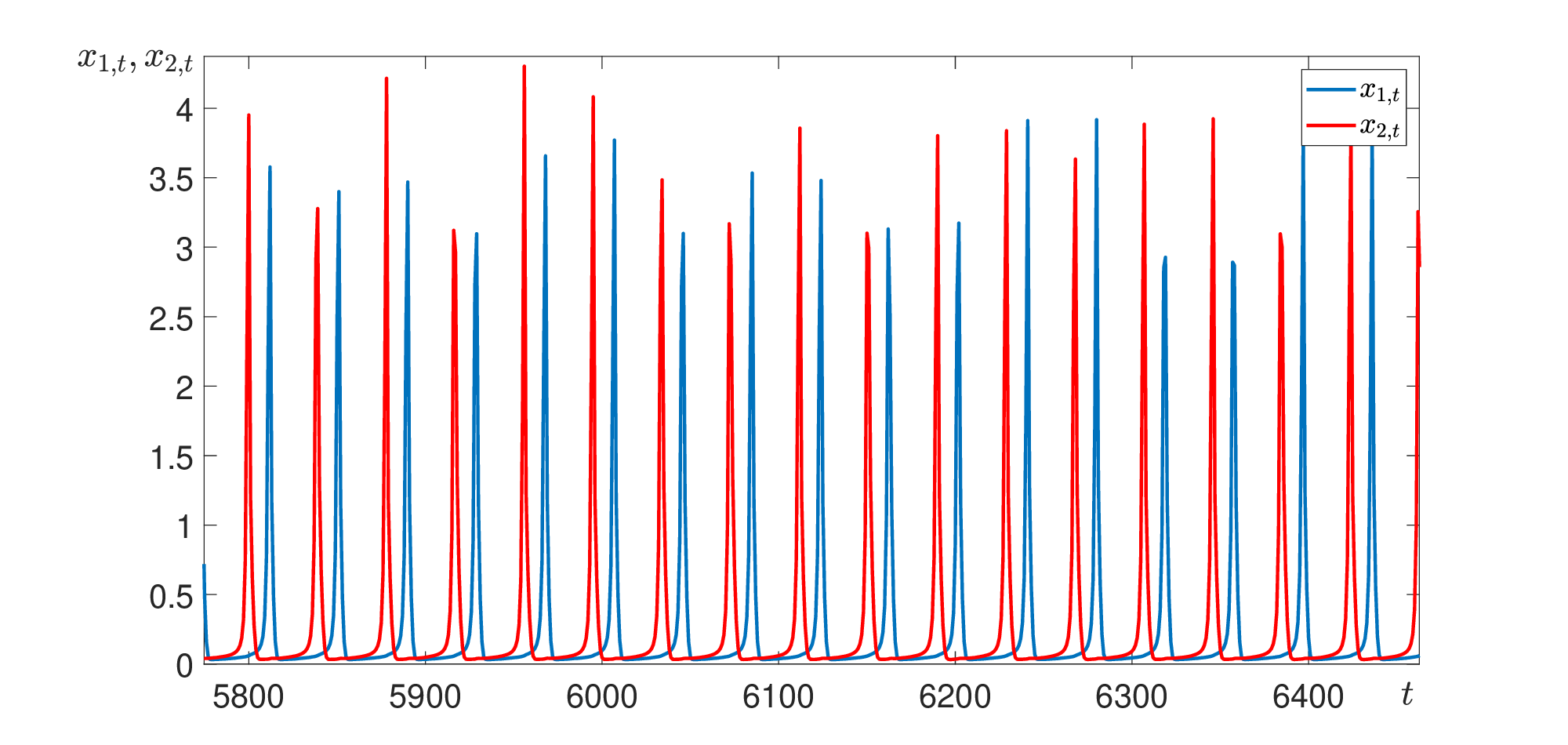}
 \caption{Time series of the two neurons are plotted for two different choices of parameters. In both figures, the model parameters and the noise intensity are the same ($a=0.89$, $b=0.35$,  $c=0.28$, $I=0.03$ and $\varepsilon=0.001$), but the coupling strength and the parameter $\Delta b$ are different: $k=0.01$ and $\Delta b=0.001$ for (a) and $k=0.001$ and  $\Delta b=-0.05$ for (b). The order parameter $R$ has been computed in both cases: in case (a) where the neurons are very well synchronized ($R=0.9729$) and in case (b) where the neurons are unsynchronized ($R=0.4434$).}
\label{figura_Rs}
\end{figure}

As an example, the time series of the variable $x$ from both neurons with different choices of parameters are plotted in Fig.~\ref{figura_Rs}.
\subsection{Excitatory coupling in the two-neuron network}\label{numericalresults2n_1}

The role played in the synchronization of the two-neuron system (Eq.~\ref{model2}) by the principal parameters of the model, is analyzed by using numerical simulations. In a first step, we consider an excitatory coupling between both neurons, and later  we proceed with the inhibitory coupling. To study the parameter dependence of $R$,  we have fixed the parameters $a=0.89$, $c=0.28$ and $I=0.03$ in (Eq.~\ref{model2}) as found in the literature (see Refs.~\cite{Chialvo95, Cazelles98, Guemez96}).

The role played by variations of the parameter $b$ has been already illustrated in Fig.~\ref{space_phase_time_series_1N_Db} and Fig.~\ref{fig_figura_ISI_b_a_89_prueba}. For our simulations, we consider three different values of the parameter $b$ corresponding to three different scenarios illustrated in Fig.~\ref{fig_figura_ISI_b_a_89_prueba}. For each value of  $b$, we analyze the effect that the noise intensity, $\varepsilon$, the coupling strength, $k$, and the mismatch in the parameter $b$, $\Delta b$, play on the order parameter $R$ and on the $ISI$ of each neuron. The value of the order parameter $R$ is obtained by computing its mean value after carrying out $50$ simulations for every point of the plane ($\varepsilon$, $\Delta b$) starting from random initial conditions.  We have considered values of $\Delta b$ up to $0.05$.

\begin{figure}[ht!]
\centering
 \includegraphics[width=\textwidth]{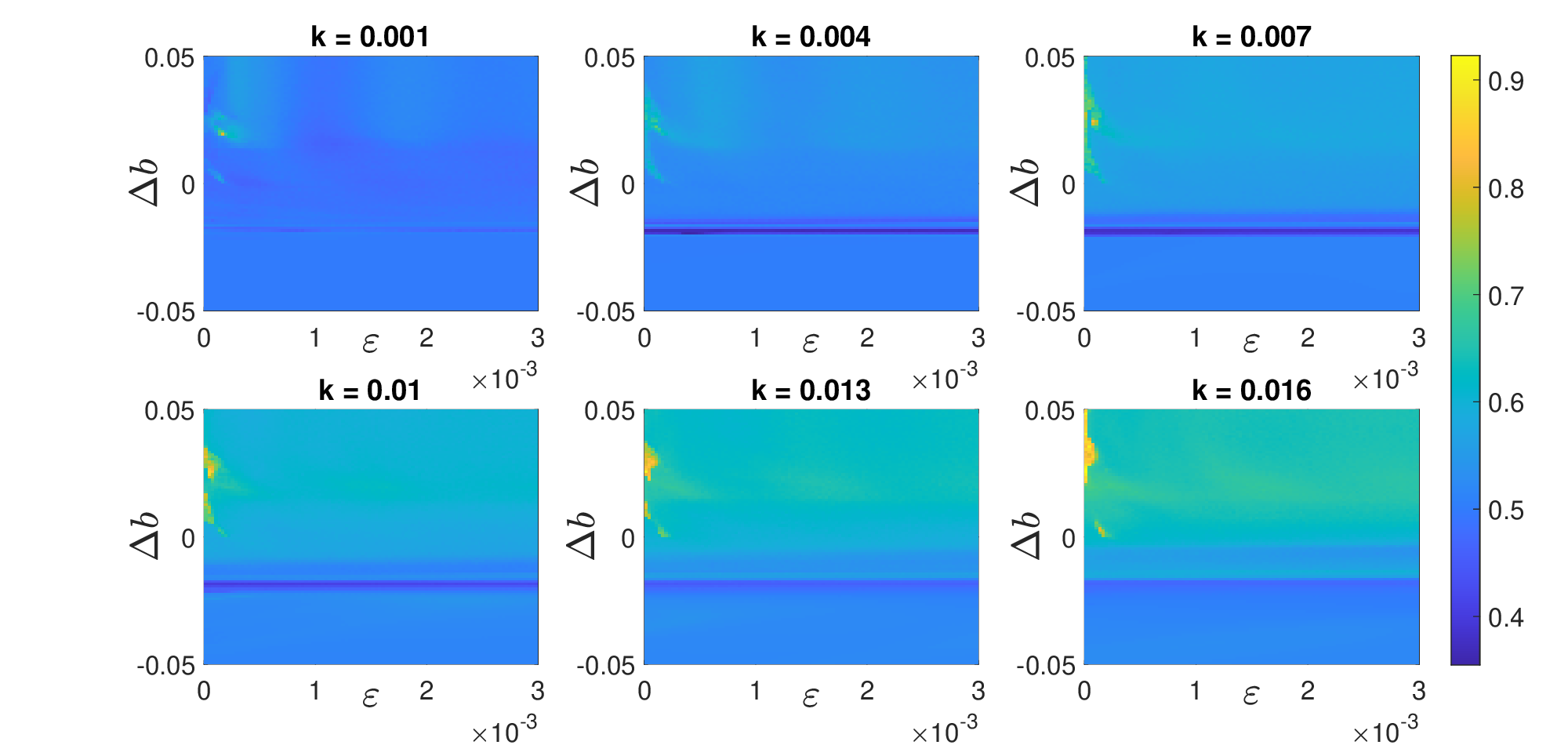}
 \caption{These figures show the order parameter $R$ with excitatory coupling between the neurons. Here, we have fixed $b_1=0.19$, $a=0.89$, $c=0.28$ and $I=0.03$. Clearly, the system is desynchronized except for a small region of parameters.}
\label{fig_a_089_b_019_Todo}
\end{figure}

We begin by considering the parameter $b_1=0.19$ which corresponds to the region where the system exhibits chaotic behavior and where the parameter $ISI$ has a steep slope as a function of $b$. Observing our results shown in Fig.~\ref{fig_a_089_b_019_Todo}, we can conclude that the system is predominantly desynchronized, except in some very small regions of the parameter plane $(\varepsilon, \Delta b)$ where $R \sim 0.85$. For larger values of $k$, the problem loses interest as the system is completely synchronized.

\begin{figure}[ht!]
\centering
 \includegraphics[width=\textwidth]{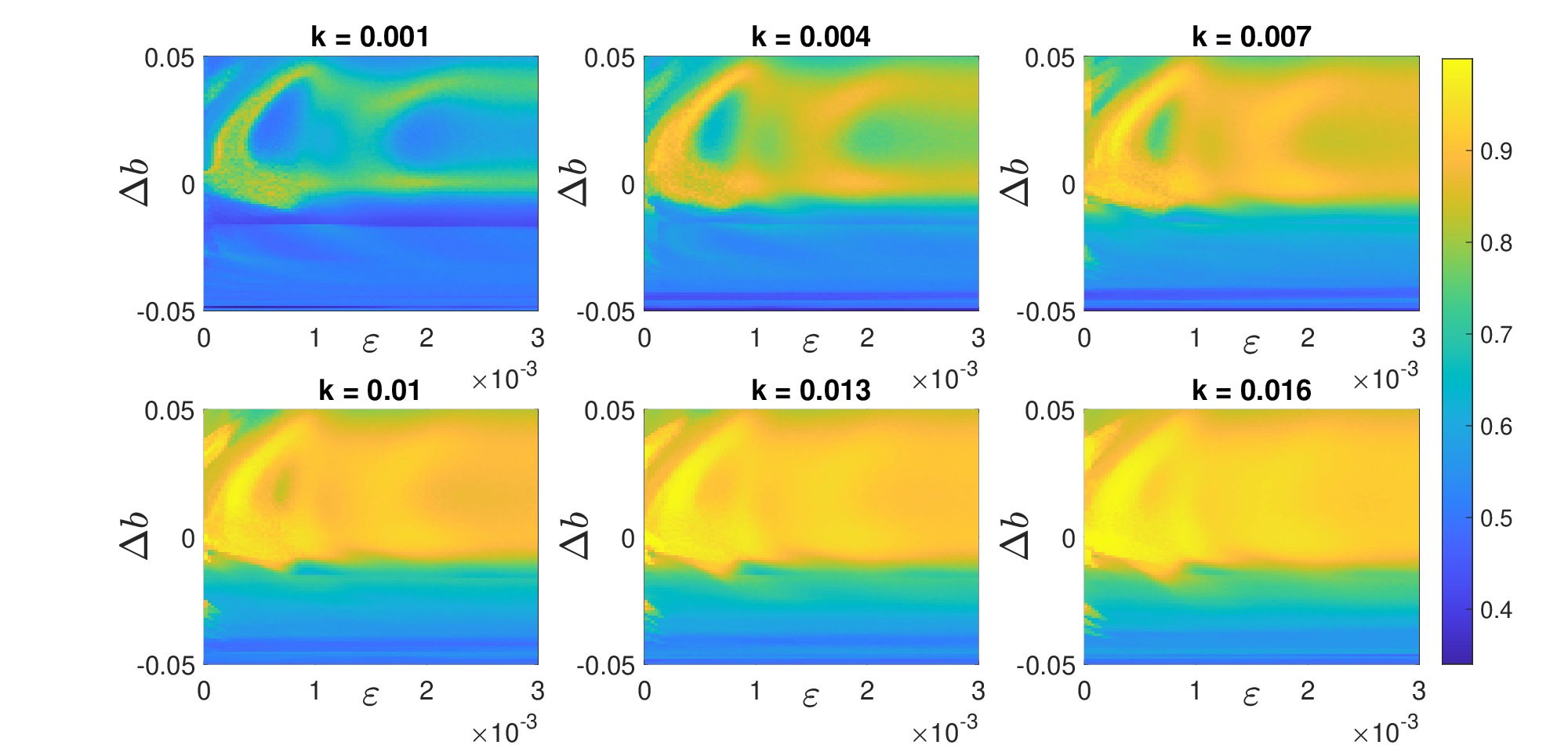}
 \caption{These figures show the order parameter $R$ with excitatory coupling between the neurons.  Here, we have fixed $b_1=0.22$, $a=0.89$, $c=0.28$ and $I=0.03$.  For increasing values of the parameter $k$, the system gets highly synchronized.}\label{fig_a_089_b_022_Todo}
\end{figure}

Now, we consider the case $b_1=0.22$ that corresponds to the region where the parameter $ISI$ is more stable. As observed in Fig.~\ref{fig_a_089_b_022_Todo}, the system is almost completely synchronized even for lower values of $k$. Moreover, for increasing values of $k$, the system remains synchronized. Additionally, in some cases an increase of the noise intensity $\varepsilon$ favors the synchronization of the system. 

One interesting observation from the figure is that the synchronization only appears when $\Delta b$ is positive (or for very small negative values). Beyond this region, the neuron $2$ has a chaotic behavior, as illustrated Fig.~\ref{Largest_Lyapunov}, preventing synchronization of the system. In cases when both neurons have Lyapunov exponents close to zero, the system can be synchronized as the noise and the coupling intensities increase.

 Of particular relevance, we note a small region where $R$ is close to $1$, centered around the point ($\varepsilon=2.5\times 10^{-4}, \Delta b=0.02$), which can be observed in all figures. As the coupling strength $k$ increases, the area of this region also increases.

\begin{figure}[ht!]
\centering
 \includegraphics[width=\textwidth]{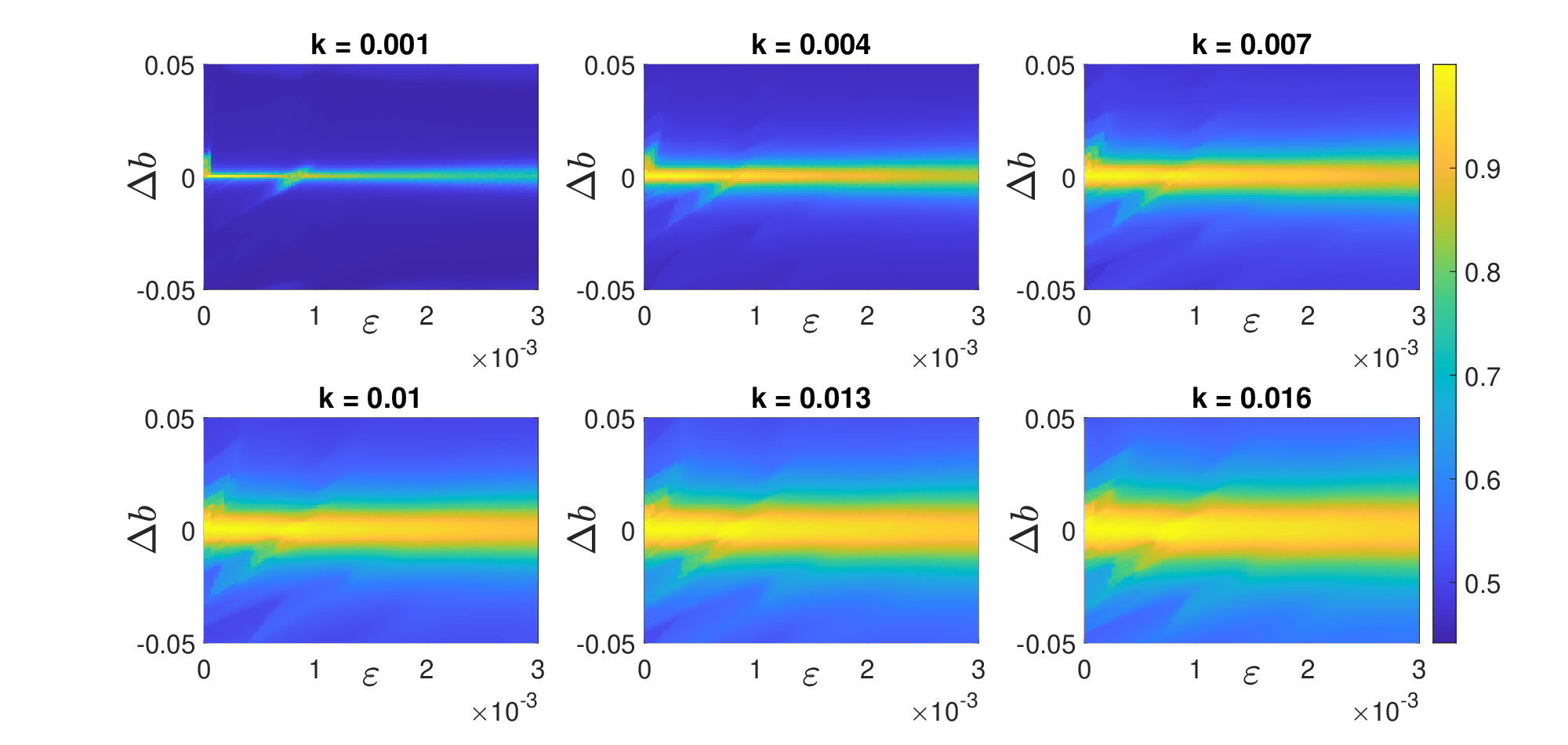}
 \caption{Figures showing the mean value of the order parameter $R$ with excitatory coupling between the neurons. Here, we have fixed $b_1=0.35$, $a=0.89$, $c=0.28$ and $I=0.03$.  We can observe that the synchronization becomes high in the region in which the parameter mismatch, $\Delta b$ is close to zero.}\label{fig_a_089_b_035_Todo}
\end{figure}

Finally, we analyze the case $b_1=0.35$, as shown in Fig.~\ref{fig_a_089_b_035_Todo}. Our results indicate that synchronization occurs when the neurons are very similar, i.e.,  $\Delta b \sim 0$. A particularly interesting phenomenon here is that the synchronization is localized \textcolor{red}{in} a narrow range of $\Delta b$. Additionally, the noise intensity does not seem to play any significant role in synchronization, as $R$ remains constant even as $\varepsilon$ increases. 

\begin{figure}[ht!]
\centering
 \includegraphics[width=\textwidth]{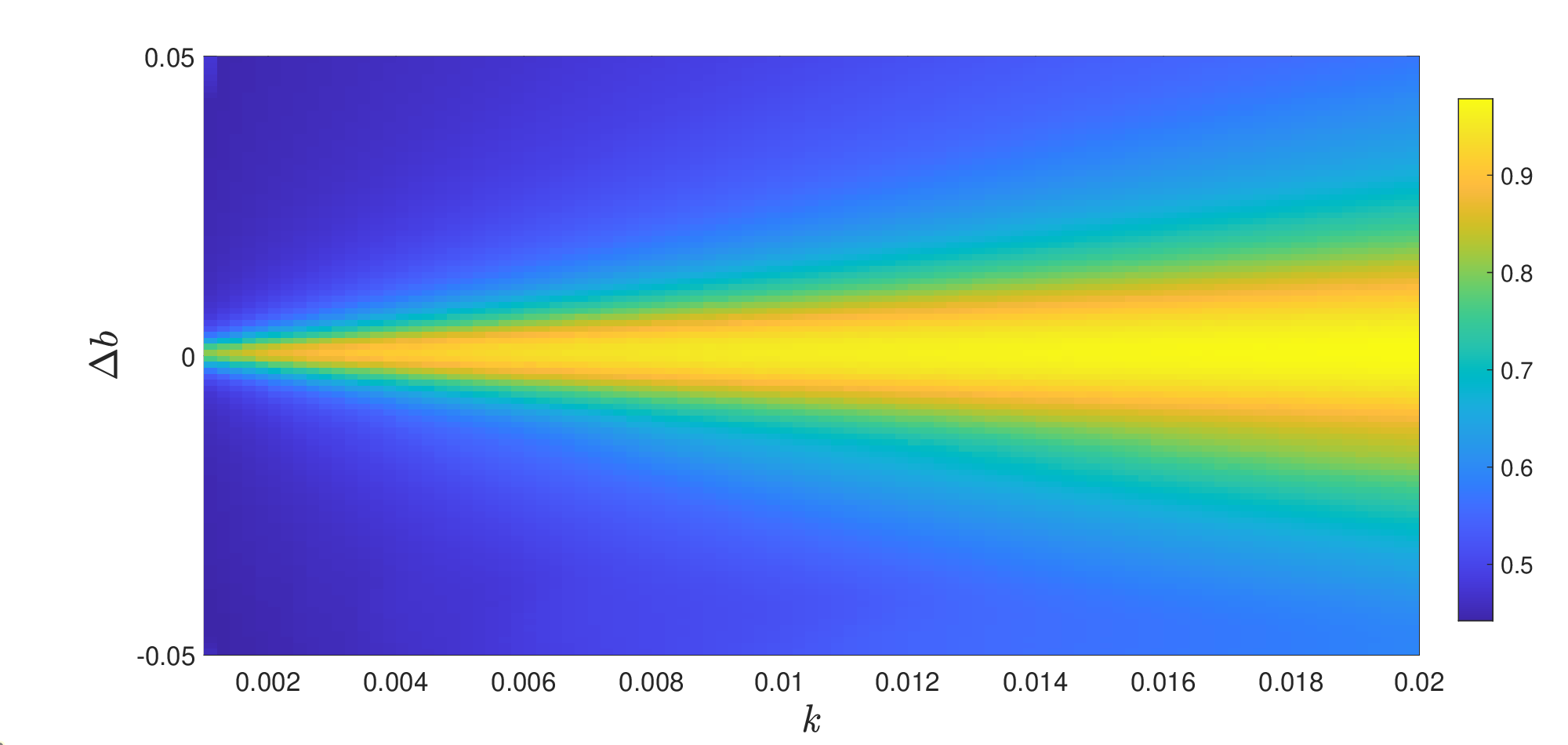}
 \caption{The figure shows how the synchronization area clearly increases with the coupling strength $k$. We have used a noise intensity $\varepsilon=0.015$, and the rest of parameters are $b_1=0.35$, $a=0.89$, $c=0.28$ and $I=0.03$. }\label{a_089_b_035_I_003_R_k_Db}
\end{figure}

It is note worthy that increasing values of the coupling strength $k$, allows neurons with quite different values of $b$ to become completely synchronized. For a more detailed analysis of this effect, we fix the noise intensity $\varepsilon=0.015$  and obtain the value of $R$ for every point of the plane ($k$, $\Delta b$). This result is shown in Fig.~\ref{a_089_b_035_I_003_R_k_Db}, where the synchronization area clearly increases with the coupling strength $k$.

The parameter $ISI$ holds significant importance in the study of the neuron networks as it is closely related with the neural coding. Here, we define $ISI_1$ as the mean value of the $ISI$ of neuron $1$ for every point of the parameter plane ($\varepsilon$, $\Delta b$) and $ISI_2$ the corresponding one for neuron $2$. To analyze the relationship between the $ISI$ of each neuron, we study the magnitude $\Delta ISI=ISI_1 - ISI_2$.

\begin{figure}[ht!]
\centering
 \includegraphics[width=\textwidth]{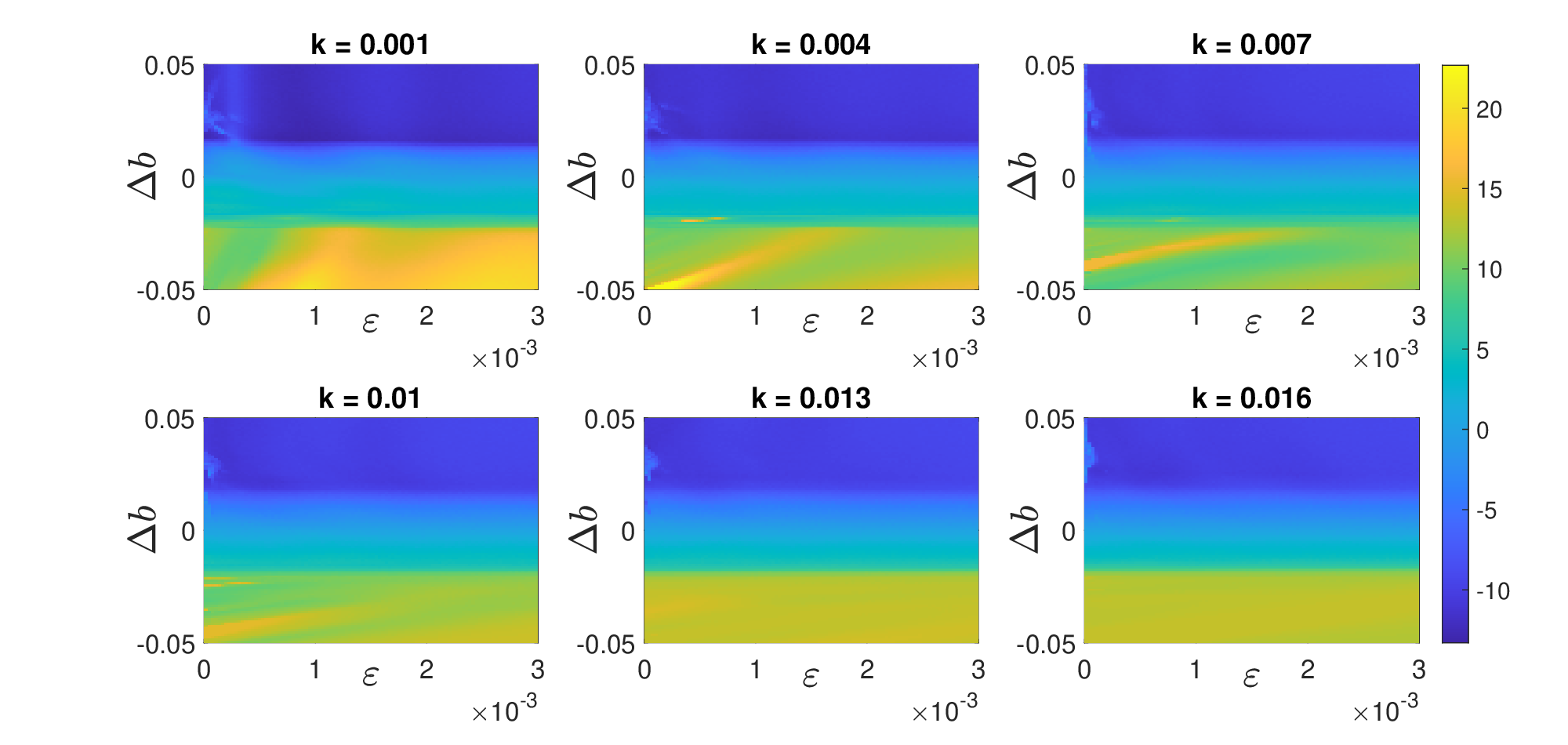}
 \caption{Representation of $\Delta ISI$ with excitatory coupling between the neurons.  We fix \textcolor{red}{the parameters as in Fig.~\ref{fig_a_089_b_019_Todo}: $b_1=0.19$, $a=0.89$, $c=0.28$ and $I=0.03$.}}
 \label{fig_a_089_b_019_D_ISI}
\end{figure}

Our strategy involves computing the  $\Delta ISI$ for the three different values of $b$ as explained in our previous simulations of $R$. The value of the $\Delta ISI$ is obtained for every point of the parameter plane ($\varepsilon$, $\Delta b$) as the mean value of $50$ simulations, each one starting from random initial conditions. We begin with $b_1=0.19$, and  our results show that $\Delta ISI$ takes values close to zero, when $\Delta b \sim 0$ (See Fig.~\ref{fig_a_089_b_019_D_ISI}). That means that both neurons have very similar spiking frequencies, even though they are not synchronized.

\begin{figure}[ht!]
\centering
 \includegraphics[width=\textwidth]{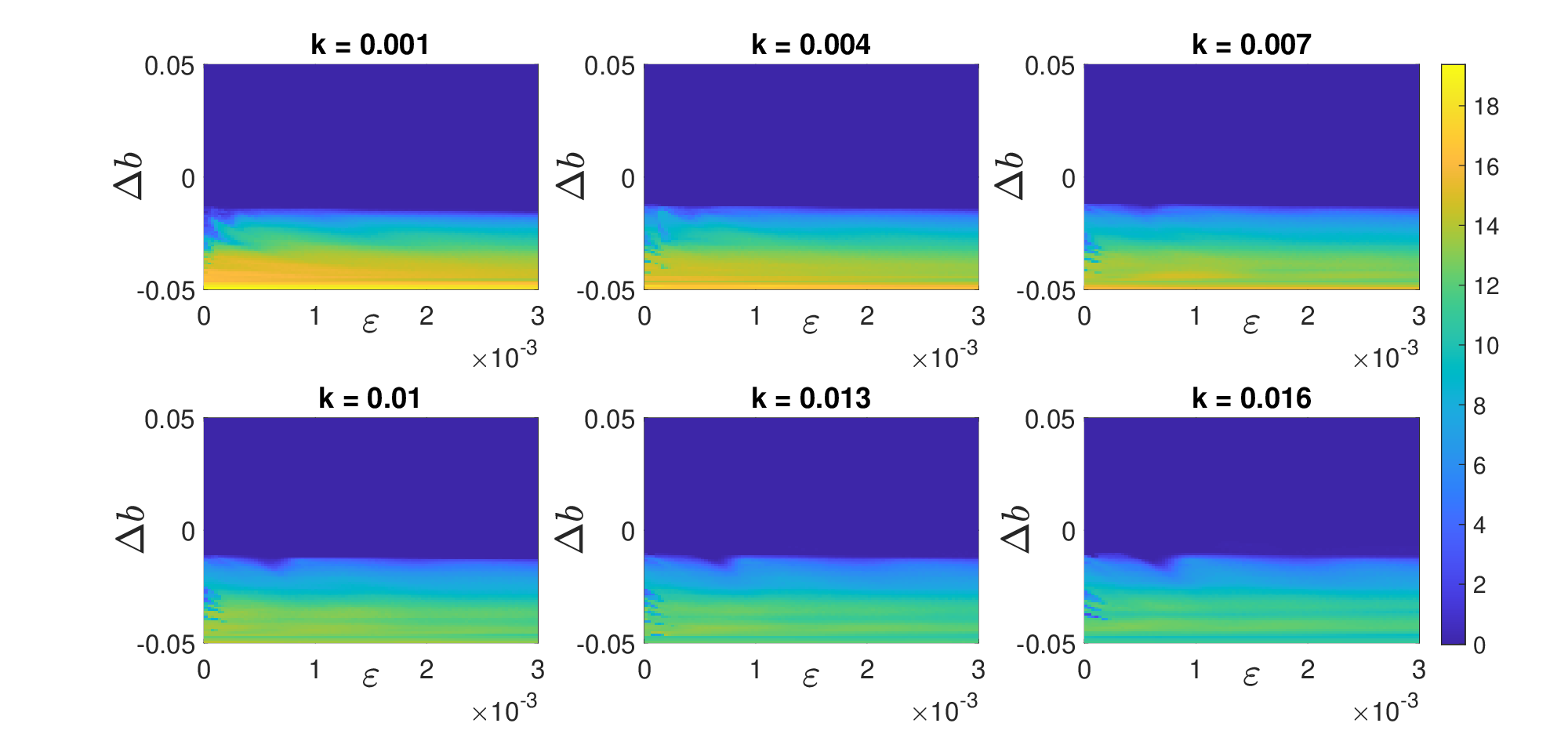}
 \caption{Representation of $\Delta ISI$ with excitatory coupling between the neurons. We fix \textcolor{red}{the parameters as in Fig.~\ref{fig_a_089_b_022_Todo}: $b_1=0.22$, $a=0.89$, $c=0.28$ and $I=0.03$.}}
\label{fig_a_089_b_022_D_ISI}
\end{figure}

Now, we consider $b_1=0.22$, and the results are shown in Fig.~\ref{fig_a_089_b_022_D_ISI} ($b_1=0.22$), where we can clearly distinguish two regions. There is a critical value $\Delta b_c=-0.012$ such that one region corresponds to $\Delta b>\Delta b_c$, where $\Delta ISI$ is close to zero, indicating that both neurons have a similar spiking frequency.  The other region corresponds to $\Delta b < \Delta b_c$, where the firing frequency is clearly different. All these results align with values of $R$ shown in Fig.~\ref{fig_a_089_b_022_Todo}.

\begin{figure}[ht!]
\centering
 \includegraphics[width=\textwidth]{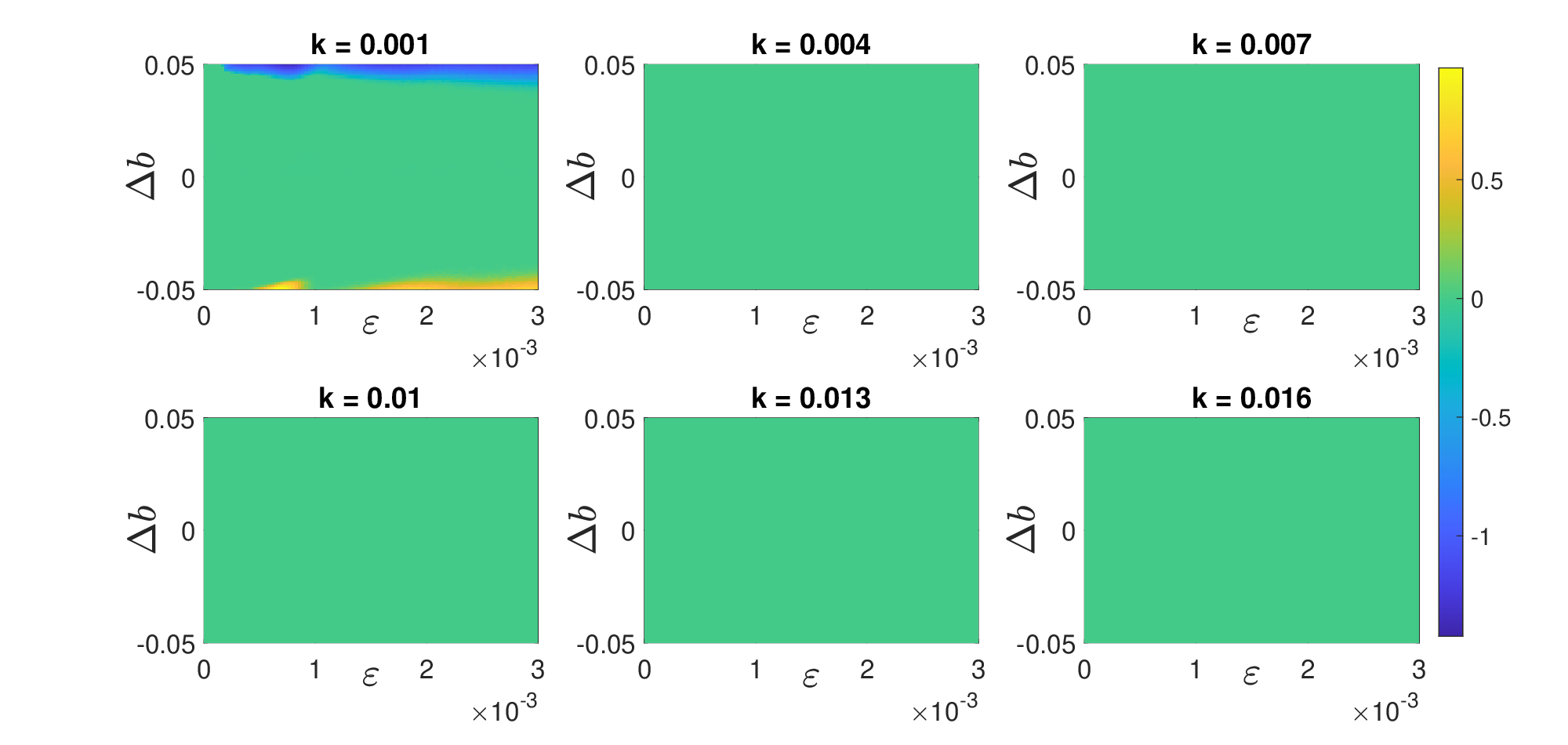}
 \caption{Representation of $\Delta ISI$ with excitatory coupling between the neurons.  We fix \textcolor{red}{the parameters as in Fig.~\ref{fig_a_089_b_035_Todo}: $b_1=0.35$, $a=0.89$, $c=0.28$ and $I=0.03$.}}
 \label{fig_a_089_b_035_D_ISI}
\end{figure}

Finally, we take $b_1=0.35$, where as observed in Fig.~\ref{fig_a_089_b_035_D_ISI},  the parameter $\Delta ISI$ is close to zero for almost every value of the coupling strength. This indicates that both neurons have the same spiking frequency. Furthermore, it suggests that the system is at least synchronized in phase, even in cases where $R$ took low values, as shown in Fig.~\ref{fig_a_089_b_035_Todo}.

\subsection{Inhibitory coupling in the two-neuron network}\label{numericalresults2n_2}

Now, we consider the two-neuron system with an inhibitory coupling.

Similar to the previous case with an excitatory coupling, our strategy consists in computing $R$ and $ISI$ for the three different values of $b$ described earlier.

\begin{figure}[ht!]
\centering
 \includegraphics[width=\textwidth]{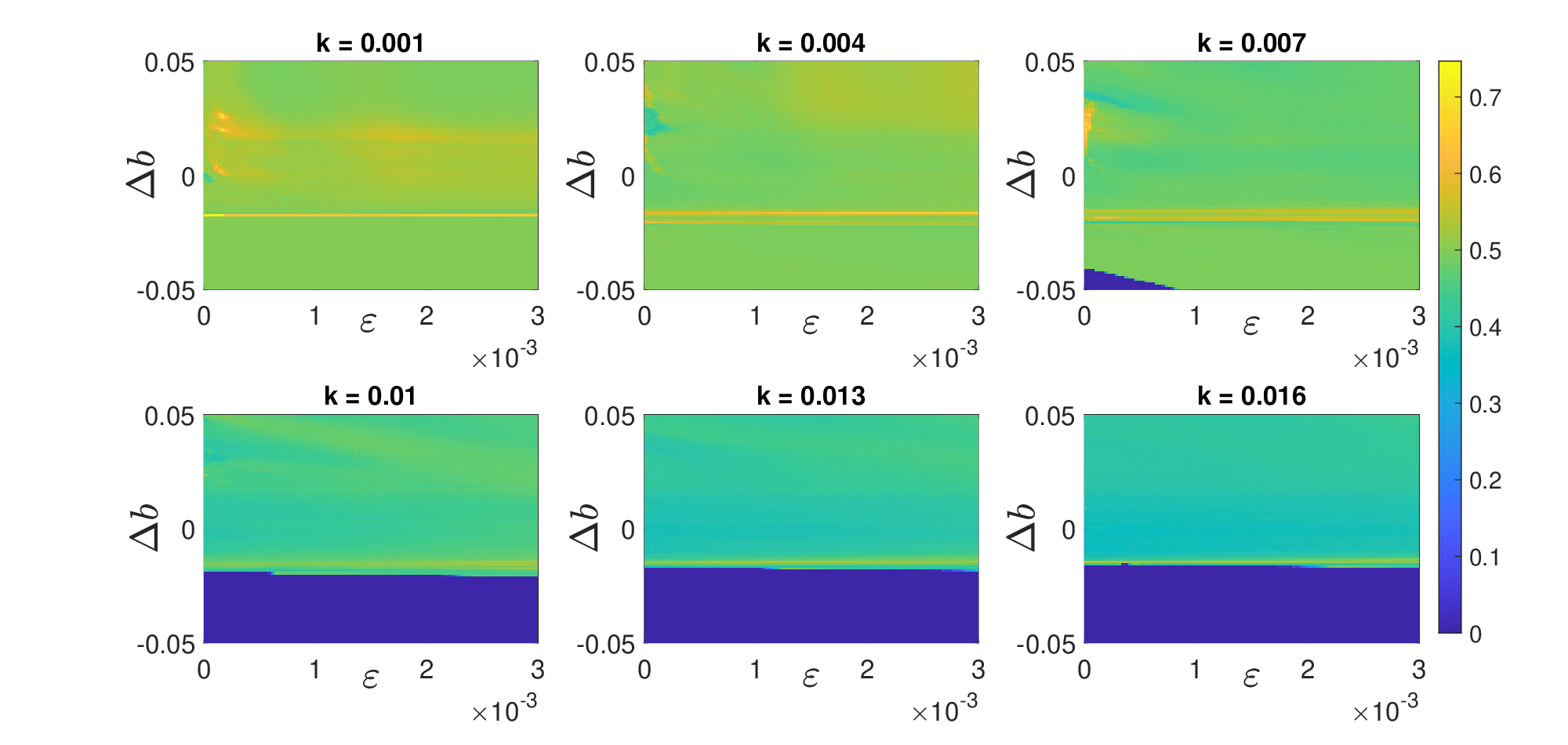}
 \caption{Plot of parameter $R$ with inhibitory coupling between the neurons. We fix $b_1=0.19$, $a=0.89$, $c=0.28$ and $I=0.03$. Here, the better synchronization, $R \simeq 0.5$, occurs for positive values of the parameter mismatch $\Delta b$.}\label{fig_a_089_b_019_Todo_Inhi}
\end{figure}

For the case $b_1=0.19$, our results, Fig.~\ref{fig_a_089_b_019_Todo_Inhi}, reveals that a region where $R$ is almost zero. This region increases as the coupling strength increases. This situation arises because the inhibitory interaction causes one of the neurons to cease its spiking behavior entering a quiescent regime.

\begin{figure}[ht!]
\centering
 \includegraphics[width=\textwidth]{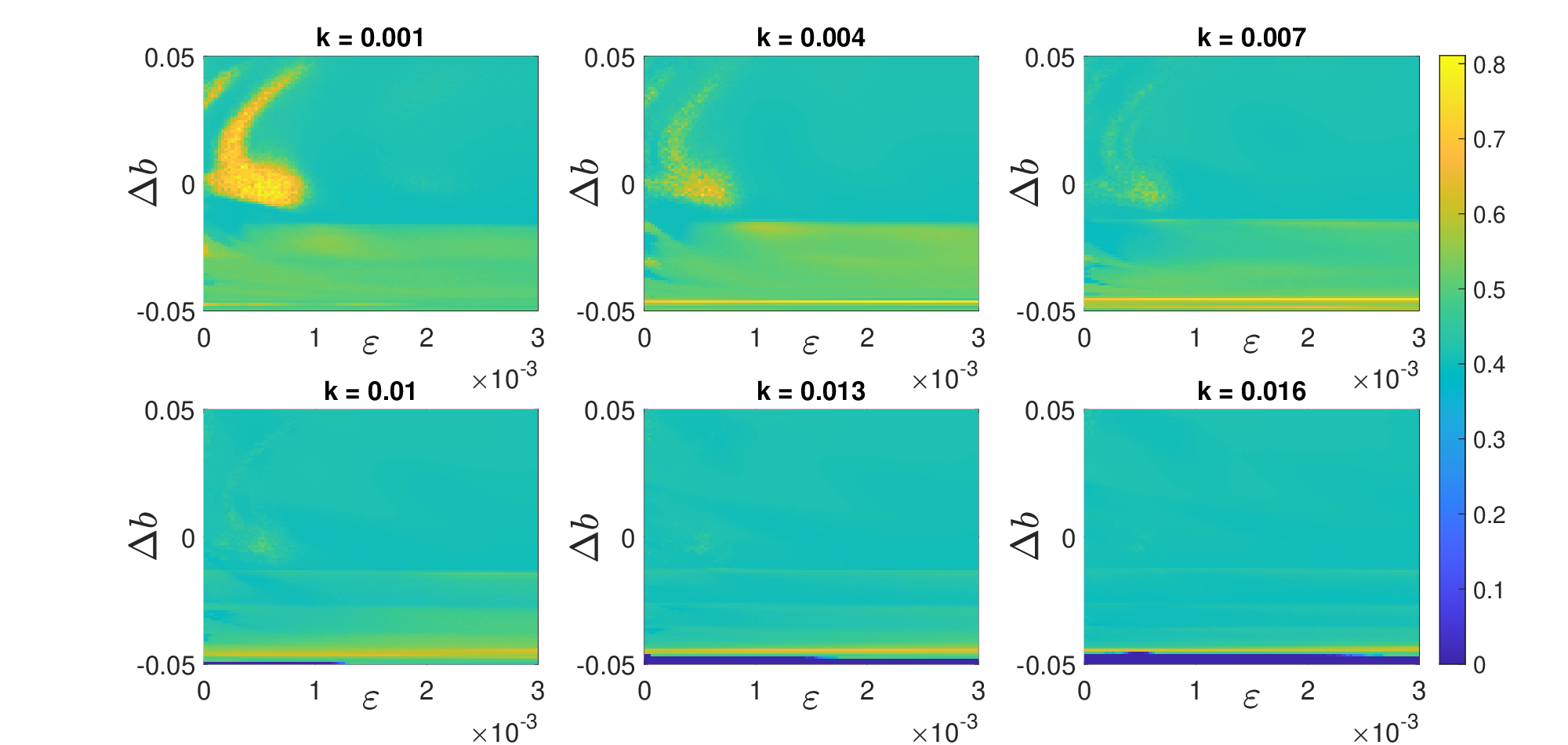}
 \caption{Plot of parameter $R$ with inhibitory coupling between the neurons. We fix $b_1=0.22$, $a=0.89$, $c=0.28$ and $I=0.03$. In this case, the highest values of $R$ are obtained for low values of both $k$ and $\Delta b$.}
 \label{fig_a_089_b_022_Todo_Inhi}
\end{figure}

Now, we consider $b_1=0.22$, and our results shown in Fig.~\ref{fig_a_089_b_022_Todo_Inhi} ($b_1=0.22$), indicate that there is a small island where the parameter $R$ takes its highest values. This area is also observed in its corresponding figure with excitatory coupling, Fig.~\ref{fig_a_089_b_019_Todo}, but with higher values of $R$. As the coupling strength increases, this area with high synchronization disappears while in the excitatory coupling this area increases. Finally, the parameter $R$ takes an almost constant value $R \sim 0.4$ for every point of the parameter plane ($\varepsilon, \Delta b$). For the highest values of the coupling, it can be observed that the parameter $R$ reaches values close to zero because as we previously explained one of the neurons changes its behavior from spiking to a quiescent state.

\begin{figure}[ht!]
\centering
 \includegraphics[width=\textwidth]{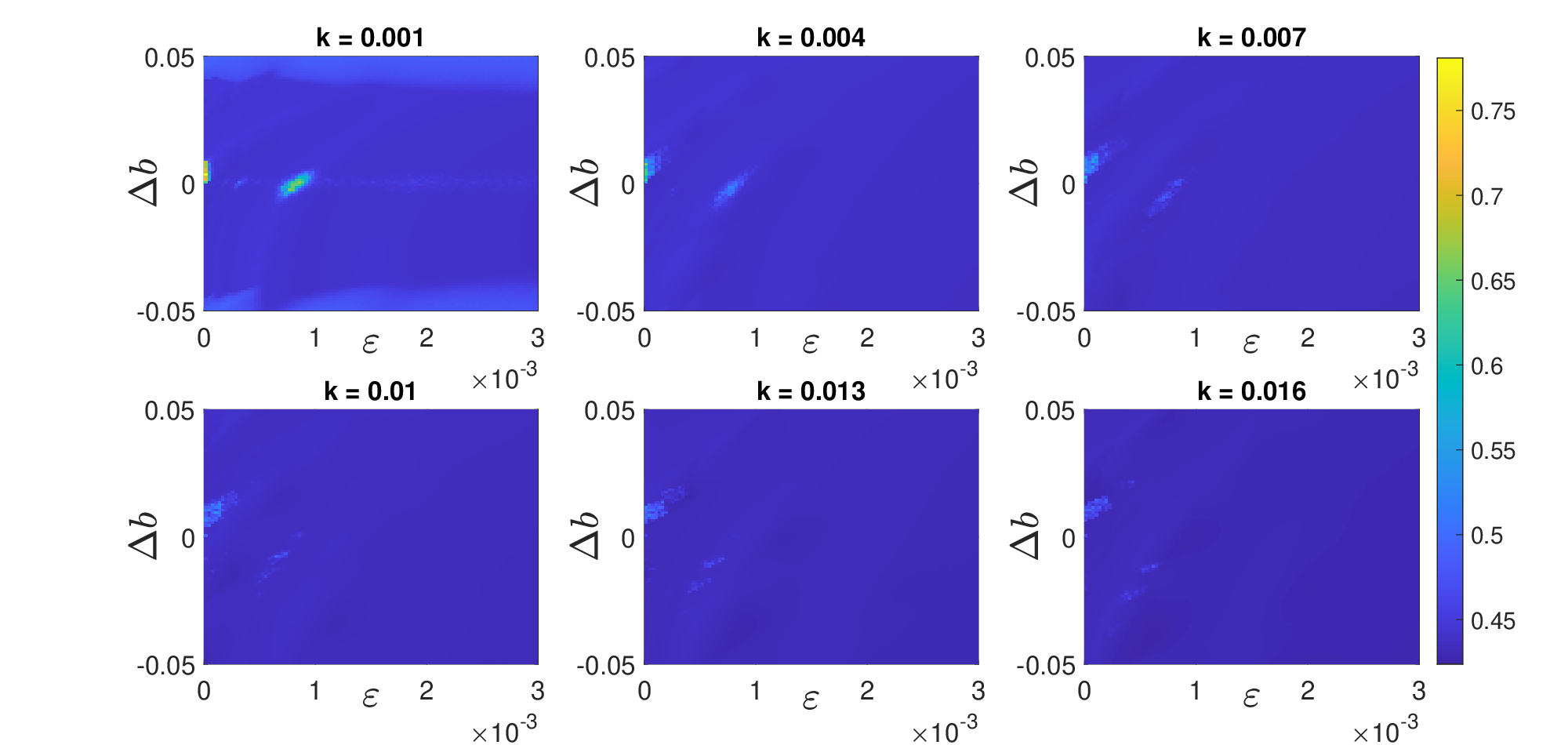}
 \caption{Plot of parameter $R$ with inhibitory coupling between the neurons. We fix $b_1=0.35$, $a=0.89$, $c=0.28$ and $I=0.03$. Here, the value of is almost constant, $R \simeq 0.5$, and it does not depend of parameter $k$.}
 \label{fig_a_089_b_035_Todo_Inhi}
\end{figure}

Finally, for $b_1=0.35$ the system is completely desynchronized, and the order parameter $R$ takes values below $0.4$ for almost all the coupling strengths and for all the points of the parameter plane ($\varepsilon, \Delta b$), see Fig.~\ref{fig_a_089_b_035_Todo_Inhi}.

Now, we proceed to analyze the effect of $\Delta b$ and $\varepsilon$ on the firing frequency of the neurons for the inhibitory case.

\begin{figure}[ht!]
\centering
 \includegraphics[width=\textwidth]{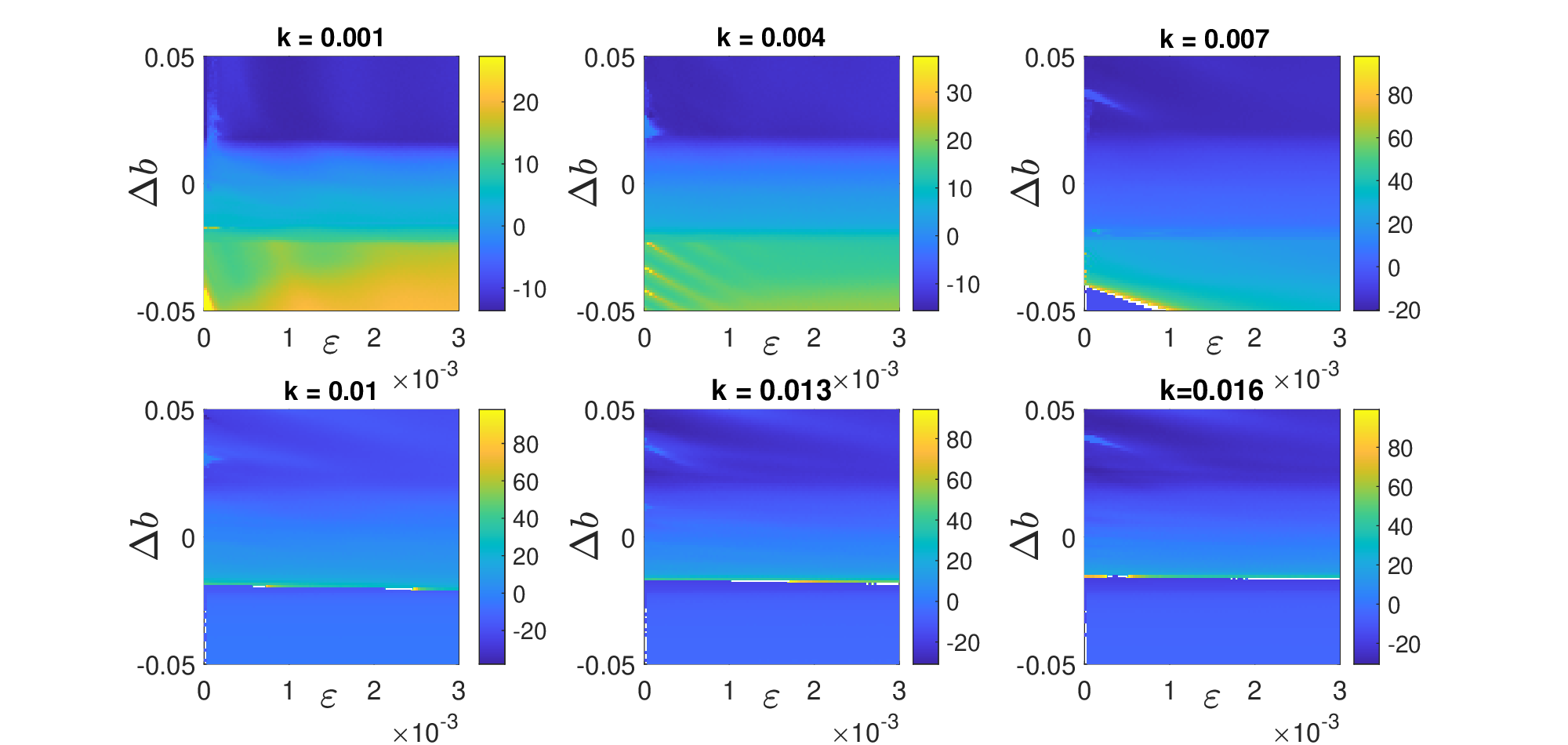}
 \caption{Representation of $\Delta ISI$ with inhibitory coupling between the neurons. We fix \textcolor{red}{the parameters as in Fig.~\ref{fig_a_089_b_019_Todo_Inhi}: $b_1=0.19$, $a=0.89$, $c=0.28$ and $I=0.03$.}}
 \label{fig_a_089_b_019_D_ISI_Inhi_2}
\end{figure}

For $b_1=0.19$, our results are shown in Fig.~\ref{fig_a_089_b_019_D_ISI_Inhi_2}, where we observe that the parameter $\Delta ISI$ only takes values close to zero in a very narrow region for positive values of $\Delta b$. This region, where both neurons have the same spiking frequency, remains quite similar for all the values of the noise intensity $\varepsilon$ and increases as the coupling strength increases. From this analysis, we can conclude that when the parameter $b$ of both neurons are in the first region studied ($b<0.20$), the neurons can only be synchronized in frequency when they are very similar or when the coupling is very strong.  Beyond of this region, their firing frequencies are quite different.

\begin{figure}[ht!]
\centering
 \includegraphics[width=\textwidth]{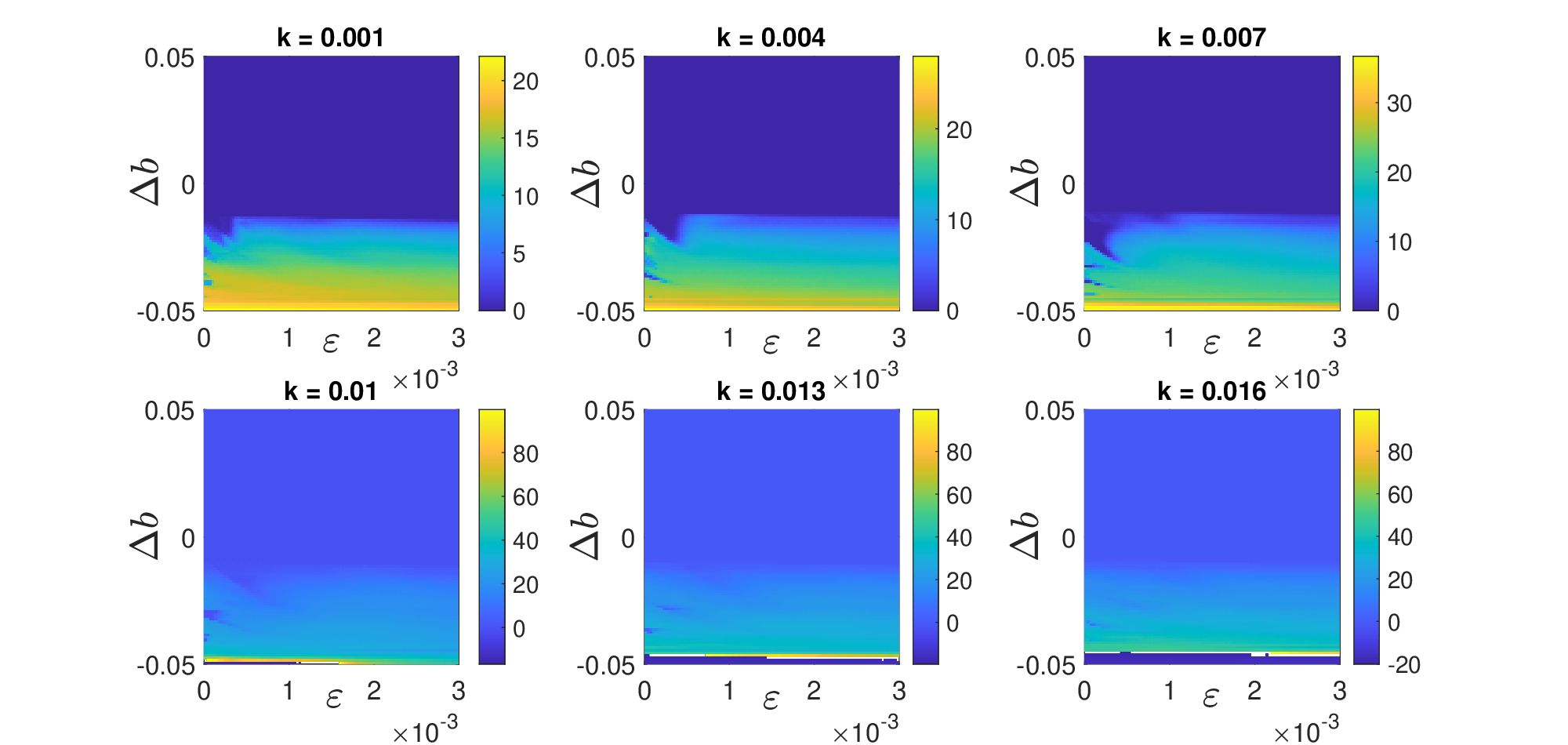}
 \caption{Representation of $\Delta ISI$ with inhibitory coupling between the neurons. We fix \textcolor{red}{the parameters as in Fig.~\ref{fig_a_089_b_022_Todo_Inhi}: $b_1=0.22$, $a=0.89$, $c=0.28$ and $I=0.03$.}}
 \label{fig_a_089_b_022_D_ISI_Inhi_2}
\end{figure}

Now, we take the value $b_1=0.22$, and the results are depicted in  Fig.~\ref{fig_a_089_b_022_D_ISI_Inhi_2}, where $\Delta ISI$ is almost zero for positive values of $\Delta b$. Additionally, this trend is also observable for small negative values of the $\Delta b$. In these cases, the neurons exhibit very similar spiking frequencies, indicating that both neurons are synchronized in phase but are not globally synchronized. However, when $\Delta b$ is negative, $\Delta ISI$ increases significantly, suggesting a substantial difference in the spiking frequencies of the neurons. This indicates that the system is neither synchronized nor in phase.

\begin{figure}[ht!]
\centering
 \includegraphics[width=\textwidth]{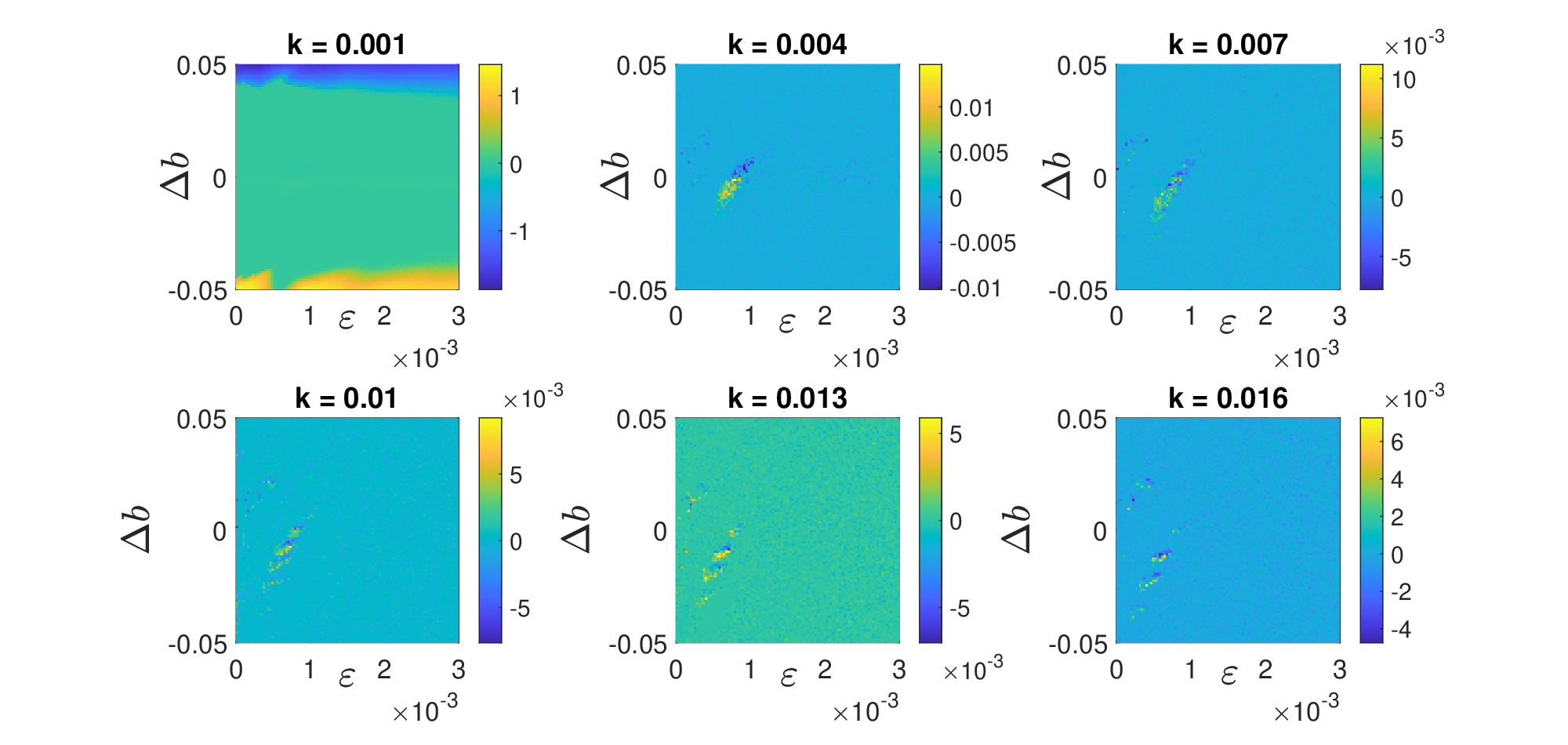}
 \caption{Representation of $\Delta ISI$ with inhibitory coupling between the neurons. We fix \textcolor{red}{the parameters as in Fig.~\ref{fig_a_089_b_035_Todo_Inhi}: $b_1=0.35$, $a=0.89$, $c=0.28$ and $I=0.03$.}}
 \label{fig_a_089_b_035_D_ISI_Inhi}
\end{figure}

Finally, for $b_1=0.35$ it can be observed in Fig.~\ref{fig_a_089_b_035_D_ISI_Inhi}, that the parameter $\Delta ISI$ is almost zero for almost every value of the coupling strength $k$. Therefore, we can conclude that here both neurons have the same firing frequency, so that both neurons are only synchronized in phase.

\section{Conclusions}
 \label{conclusions}

We have conducted a study on the synchronization of two electrically coupled non-identical neurons as a function of the coupling strength between the neurons, the noise intensity and the difference between neurons by introducing a mismatch in one of the parameters of the stochastic map-based Chialvo model.

Our strategy has consisted in considering three different values of parameter $b$ each representing diverse behaviors of the neuron. Our aim was to elucidate the conditions under which two neurons with very different behavior could be synchronized.

As expected, the type of coupling between neurons plays a fundamental role. Specifically, when the coupling is excitatory, the system tends to become  completely synchronized. Conversely, when the coupling is inhibitory the system tends to become highly desynchronized.

Our results clearly show that the mismatch parameter $\Delta b$ plays a key role.  In particular,  when both neurons take a value of the parameter $b\sim 0.19$, what means that both neurons have a chaotic behavior, only in a very small region with low coupling strength the system can be synchronized. On the contrary when $b\sim 0.22$, we find a critical value $\Delta b_c$ such that for $\Delta b_c \le \Delta b$ the synchronization of the neurons is enhanced as the coupling strength and the noise intensity increases. Finally, when both neurons has a stable cycle behavior, $b\sim 0.35$, the synchronization is only possible when they are quite similar ($\Delta b \sim 0$), otherwise the synchronization decreases abruptly  as $\Delta b$ increases. Besides, the range of $\Delta b$ for which the system has a good synchronization increases with the coupling strength.

An analysis of the inter-spike interval ($ISI$) of the neurons has been conducted. For excitatory coupling, the parameter $\Delta ISI$ remains constant regardless of the noise intensity, except for very low values of the coupling strength when $b\sim 0.19$. However, for $b\sim 0.22$, the critical value $\Delta b_c$ mentioned earlier leads to the emerge of two distinct regimes. In one of these regions,  $\Delta b>\Delta b_c$ both neurons exhibit a similar spiking frequency. In the other region, $\Delta b < \Delta b_c$, the firing frequency increases as the parameter $\Delta b$ increases. When $b\sim 0.35$, the information provided by the parameter $\Delta ISI$ indicates that both neurons have the same spiking frequency, suggesting that the system is at least synchronized in phase.

When the coupling is inhibitory the system becomes highly desynchronized. Notably, the system is only synchronized for low values of the coupling strength when $b\sim 0.19$ and $b\sim 0.22$. As in the excitatory case, it can be observed that neurons have similar firing frequency only when both neurons are similar. Nevertheless, in the inhibitory case when the neurons are different, the value of $\Delta ISI$ is much higher than in the excitatory case. Lastly, for $b \sim 0.35$ the value of $\Delta ISI$ is very close to zero what means that despite the neurons being highly desynchronized, their firing frequency is the same, so that we can assure that the neurons are at least synchronized in phase.

\noindent{\bf{Acknowledgments}}

JU, JMS and MAFS acknowledge financial support from the Spanish State Research Agency (AEI) and the European Regional Development Fund (ERDF, EU) under Project No.~PID2019-105554GB-I00 (MCIN/AEI/10.13039/501100011033).

\bibliography{Bib_Neuron_Map}

\begin{thebibliography}{25}
\expandafter\ifx\csname natexlab\endcsname\relax\def\natexlab#1{#1}\fi
\expandafter\ifx\csname bibnamefont\endcsname\relax
  \def\bibnamefont#1{#1}\fi
\expandafter\ifx\csname bibfnamefont\endcsname\relax
  \def\bibfnamefont#1{#1}\fi
\expandafter\ifx\csname citenamefont\endcsname\relax
  \def\citenamefont#1{#1}\fi
\expandafter\ifx\csname url\endcsname\relax
  \def\url#1{\texttt{#1}}\fi
\expandafter\ifx\csname urlprefix\endcsname\relax\def\urlprefix{URL }\fi
\providecommand{\bibinfo}[2]{#2}
\providecommand{\eprint}[2][]{\url{#2}}

\bibitem[{\citenamefont{Hodgkin and Huxley}(1952{\natexlab{a}})}]{Hodgkin1952}
\bibinfo{author}{\bibfnamefont{A.~L.} \bibnamefont{Hodgkin}} \bibnamefont{and}
  \bibinfo{author}{\bibfnamefont{A.~F.} \bibnamefont{Huxley}},
  \bibinfo{journal}{The Journal of Physiology} \textbf{\bibinfo{volume}{116}},
  \bibinfo{pages}{449} (\bibinfo{year}{1952}{\natexlab{a}}).

\bibitem[{\citenamefont{Hodgkin and
  Huxley}(1952{\natexlab{b}})}]{Hodgkin1952_2}
\bibinfo{author}{\bibfnamefont{A.~L.} \bibnamefont{Hodgkin}} \bibnamefont{and}
  \bibinfo{author}{\bibfnamefont{A.~F.} \bibnamefont{Huxley}},
  \bibinfo{journal}{The Journal of Physiology} \textbf{\bibinfo{volume}{117}},
  \bibinfo{pages}{500} (\bibinfo{year}{1952}{\natexlab{b}}).

\bibitem[{\citenamefont{FitzHugh}(1961)}]{FitzHugh1961}
\bibinfo{author}{\bibfnamefont{R.}~\bibnamefont{FitzHugh}},
  \bibinfo{journal}{Biophys. J.} \textbf{\bibinfo{volume}{1}},
  \bibinfo{pages}{445} (\bibinfo{year}{1961}).

\bibitem[{\citenamefont{Morris and Lecar}(1981)}]{morris:lecar}
\bibinfo{author}{\bibfnamefont{C.}~\bibnamefont{Morris}} \bibnamefont{and}
  \bibinfo{author}{\bibfnamefont{H.}~\bibnamefont{Lecar}},
  \bibinfo{journal}{Biophys. J.} \textbf{\bibinfo{volume}{35}},
  \bibinfo{pages}{193} (\bibinfo{year}{1981}).

\bibitem[{\citenamefont{Hindmarsh and Rose}(1982)}]{Hindmarsh1982}
\bibinfo{author}{\bibfnamefont{J.~L.} \bibnamefont{Hindmarsh}}
  \bibnamefont{and} \bibinfo{author}{\bibfnamefont{R.~M.} \bibnamefont{Rose}},
  \bibinfo{journal}{Nature} \textbf{\bibinfo{volume}{296}},
  \bibinfo{pages}{162} (\bibinfo{year}{1982}).

\bibitem[{\citenamefont{Shilnikov and Kolomiets}(2008)}]{Shilnikov2008}
\bibinfo{author}{\bibfnamefont{A.}~\bibnamefont{Shilnikov}} \bibnamefont{and}
  \bibinfo{author}{\bibfnamefont{M.}~\bibnamefont{Kolomiets}},
  \bibinfo{journal}{International Journal of Bifurcation and Chaos}
  \textbf{\bibinfo{volume}{18}}, \bibinfo{pages}{2141} (\bibinfo{year}{2008}).

\bibitem[{\citenamefont{Rulkov}(2001)}]{Rulkov01}
\bibinfo{author}{\bibfnamefont{N.~F.} \bibnamefont{Rulkov}},
  \bibinfo{journal}{Phys. Rev. Lett.} \textbf{\bibinfo{volume}{86}},
  \bibinfo{pages}{183} (\bibinfo{year}{2001}).

\bibitem[{\citenamefont{Ibarz et~al.}(2011)\citenamefont{Ibarz, Casado, and
  Sanju\'an}}]{Ibarz2011}
\bibinfo{author}{\bibfnamefont{B.}~\bibnamefont{Ibarz}},
  \bibinfo{author}{\bibfnamefont{J.~M.} \bibnamefont{Casado}},
  \bibnamefont{and} \bibinfo{author}{\bibfnamefont{M.~A.~F.}
  \bibnamefont{Sanju\'an}}, \bibinfo{journal}{Phys. Rep.}
  \textbf{\bibinfo{volume}{501}}, \bibinfo{pages}{1} (\bibinfo{year}{2011}).

\bibitem[{\citenamefont{Girardi-Schappo
  et~al.}(2013)\citenamefont{Girardi-Schappo, Tragtenberg, and
  Kinouchi}}]{Gir2013}
\bibinfo{author}{\bibfnamefont{M.}~\bibnamefont{Girardi-Schappo}},
  \bibinfo{author}{\bibfnamefont{M.}~\bibnamefont{Tragtenberg}},
  \bibnamefont{and} \bibinfo{author}{\bibfnamefont{O.}~\bibnamefont{Kinouchi}},
  \bibinfo{journal}{Journal of Neuroscience Methods}
  \textbf{\bibinfo{volume}{220}}, \bibinfo{pages}{116 } (\bibinfo{year}{2013}).

\bibitem[{\citenamefont{Kinouchi and Tragtenberg}(1996)}]{KT96}
\bibinfo{author}{\bibfnamefont{O.}~\bibnamefont{Kinouchi}} \bibnamefont{and}
  \bibinfo{author}{\bibfnamefont{M.~H.~R.} \bibnamefont{Tragtenberg}},
  \bibinfo{journal}{Int. J. Bifurcat. Chaos} \textbf{\bibinfo{volume}{6}},
  \bibinfo{pages}{2343} (\bibinfo{year}{1996}).

\bibitem[{\citenamefont{Chialvo}(1995)}]{Chialvo95}
\bibinfo{author}{\bibfnamefont{D.~R.} \bibnamefont{Chialvo}},
  \bibinfo{journal}{Chaos, Solitons \& Fractals} \textbf{\bibinfo{volume}{5}},
  \bibinfo{pages}{461 } (\bibinfo{year}{1995}).

\bibitem[{\citenamefont{Wang and Cao}(2018)}]{WANG2018}
\bibinfo{author}{\bibfnamefont{F.}~\bibnamefont{Wang}} \bibnamefont{and}
  \bibinfo{author}{\bibfnamefont{H.}~\bibnamefont{Cao}},
  \bibinfo{journal}{Communications in Nonlinear Science and Numerical
  Simulation} \textbf{\bibinfo{volume}{56}}, \bibinfo{pages}{481}
  (\bibinfo{year}{2018}), ISSN \bibinfo{issn}{1007-5704}.

\bibitem[{\citenamefont{Yang et~al.}(2020)\citenamefont{Yang, Xiang, Dai, Qi,
  and Dong}}]{Yang2020}
\bibinfo{author}{\bibfnamefont{Y.}~\bibnamefont{Yang}},
  \bibinfo{author}{\bibfnamefont{C.}~\bibnamefont{Xiang}},
  \bibinfo{author}{\bibfnamefont{X.}~\bibnamefont{Dai}},
  \bibinfo{author}{\bibfnamefont{L.}~\bibnamefont{Qi}}, \bibnamefont{and}
  \bibinfo{author}{\bibfnamefont{T.}~\bibnamefont{Dong}}, in
  \emph{\bibinfo{booktitle}{Advances in Neural Networks -- ISNN 2020}}
  (\bibinfo{publisher}{Springer International Publishing},
  \bibinfo{year}{2020}), pp. \bibinfo{pages}{61--73}, ISBN
  \bibinfo{isbn}{978-3-030-64221-1}.

\bibitem[{\citenamefont{Bashkirtseva et~al.}(2023)\citenamefont{Bashkirtseva,
  Ryashko, Used, Seoane, and Sanju\'an}}]{bruss:2023}
\bibinfo{author}{\bibfnamefont{I.}~\bibnamefont{Bashkirtseva}},
  \bibinfo{author}{\bibfnamefont{L.}~\bibnamefont{Ryashko}},
  \bibinfo{author}{\bibfnamefont{J.}~\bibnamefont{Used}},
  \bibinfo{author}{\bibfnamefont{J.}~\bibnamefont{Seoane}}, \bibnamefont{and}
  \bibinfo{author}{\bibfnamefont{M.~A.~F.} \bibnamefont{Sanju\'an}},
  \bibinfo{journal}{Communications in Nonlinear Science and Numerical
  Simulation} \textbf{\bibinfo{volume}{116}}, \bibinfo{pages}{1}
  (\bibinfo{year}{2023}).

\bibitem[{\citenamefont{Zambrano et~al.}(2010)\citenamefont{Zambrano,
  Mari\~{n}o, Seoane, , Sanju\'an, Euzzor, Geltrude, Meucci, and
  Arecchi}}]{zambrano:2010}
\bibinfo{author}{\bibfnamefont{S.}~\bibnamefont{Zambrano}},
  \bibinfo{author}{\bibfnamefont{I.~P.} \bibnamefont{Mari\~{n}o}},
  \bibinfo{author}{\bibfnamefont{J.~M.} \bibnamefont{Seoane}}, ,
  \bibinfo{author}{\bibfnamefont{M.~A.~F.} \bibnamefont{Sanju\'an}},
  \bibinfo{author}{\bibfnamefont{S.}~\bibnamefont{Euzzor}},
  \bibinfo{author}{\bibfnamefont{A.}~\bibnamefont{Geltrude}},
  \bibinfo{author}{\bibfnamefont{R.}~\bibnamefont{Meucci}}, \bibnamefont{and}
  \bibinfo{author}{\bibfnamefont{F.~T.} \bibnamefont{Arecchi}},
  \bibinfo{journal}{New Journal of Physics} \textbf{\bibinfo{volume}{12}},
  \bibinfo{pages}{1} (\bibinfo{year}{2010}).

\bibitem[{\citenamefont{Sriram et~al.}(2023)\citenamefont{Sriram, Mirzaei,
  Mehrabbeik, Rajagopal, Rostami, and Jafari}}]{SRIRAM2023}
\bibinfo{author}{\bibfnamefont{S.}~\bibnamefont{Sriram}},
  \bibinfo{author}{\bibfnamefont{S.}~\bibnamefont{Mirzaei}},
  \bibinfo{author}{\bibfnamefont{M.}~\bibnamefont{Mehrabbeik}},
  \bibinfo{author}{\bibfnamefont{K.}~\bibnamefont{Rajagopal}},
  \bibinfo{author}{\bibfnamefont{M.}~\bibnamefont{Rostami}}, \bibnamefont{and}
  \bibinfo{author}{\bibfnamefont{S.}~\bibnamefont{Jafari}},
  \bibinfo{journal}{Journal of Theoretical Biology}
  \textbf{\bibinfo{volume}{572}}, \bibinfo{pages}{111591}
  (\bibinfo{year}{2023}), ISSN \bibinfo{issn}{0022-5193}.

\bibitem[{\citenamefont{Muni et~al.}(2022)\citenamefont{Muni, Fatoyinbo, and
  Ghosh}}]{Muni_2022}
\bibinfo{author}{\bibfnamefont{S.~S.} \bibnamefont{Muni}},
  \bibinfo{author}{\bibfnamefont{H.~O.} \bibnamefont{Fatoyinbo}},
  \bibnamefont{and} \bibinfo{author}{\bibfnamefont{I.}~\bibnamefont{Ghosh}},
  \bibinfo{journal}{International Journal of Bifurcation and Chaos}
  \textbf{\bibinfo{volume}{32}} (\bibinfo{year}{2022}).

\bibitem[{\citenamefont{Vivekanandhan et~al.}(2023)\citenamefont{Vivekanandhan,
  Natiq, Merrikhi, Rajagopal, and Jafari}}]{Vivekanandhan_2023}
\bibinfo{author}{\bibfnamefont{G.}~\bibnamefont{Vivekanandhan}},
  \bibinfo{author}{\bibfnamefont{H.}~\bibnamefont{Natiq}},
  \bibinfo{author}{\bibfnamefont{Y.}~\bibnamefont{Merrikhi}},
  \bibinfo{author}{\bibfnamefont{K.}~\bibnamefont{Rajagopal}},
  \bibnamefont{and} \bibinfo{author}{\bibfnamefont{S.}~\bibnamefont{Jafari}},
  \bibinfo{journal}{Electronics} \textbf{\bibinfo{volume}{12}}
  (\bibinfo{year}{2023}), ISSN \bibinfo{issn}{2079-9292}.

\bibitem[{\citenamefont{Hoppensteadt}(1986)}]{Hoppensteadt86}
\bibinfo{author}{\bibfnamefont{F.~C.} \bibnamefont{Hoppensteadt}},
  \emph{\bibinfo{title}{An Introduction to the Mathematics of Neurons}}
  (\bibinfo{publisher}{Cambridge University Press},
  \bibinfo{address}{Cambridge}, \bibinfo{year}{1986}).

\bibitem[{\citenamefont{Fujii et~al.}(1996)\citenamefont{Fujii, Ito, Aidhara,
  Ichinose, and Tsukada}}]{Fujii96}
\bibinfo{author}{\bibfnamefont{H.}~\bibnamefont{Fujii}},
  \bibinfo{author}{\bibfnamefont{H.}~\bibnamefont{Ito}},
  \bibinfo{author}{\bibfnamefont{K.}~\bibnamefont{Aidhara}},
  \bibinfo{author}{\bibfnamefont{N.}~\bibnamefont{Ichinose}}, \bibnamefont{and}
  \bibinfo{author}{\bibfnamefont{M.}~\bibnamefont{Tsukada}},
  \bibinfo{journal}{Neural Networks} \textbf{\bibinfo{volume}{9}},
  \bibinfo{pages}{1303 } (\bibinfo{year}{1996}).

\bibitem[{\citenamefont{MacLeod and Laurent}(1996)}]{MacLeod96}
\bibinfo{author}{\bibfnamefont{K.}~\bibnamefont{MacLeod}} \bibnamefont{and}
  \bibinfo{author}{\bibfnamefont{G.}~\bibnamefont{Laurent}},
  \bibinfo{journal}{Science} \textbf{\bibinfo{volume}{274}},
  \bibinfo{pages}{976 } (\bibinfo{year}{1996}).

\bibitem[{\citenamefont{Tanaka et~al.}(2006)\citenamefont{Tanaka, Ibarz,
  Sanju\'{a}n, and Aihara}}]{TISA:2006}
\bibinfo{author}{\bibfnamefont{G.}~\bibnamefont{Tanaka}},
  \bibinfo{author}{\bibfnamefont{B.}~\bibnamefont{Ibarz}},
  \bibinfo{author}{\bibfnamefont{M.~A.~F.} \bibnamefont{Sanju\'{a}n}},
  \bibnamefont{and} \bibinfo{author}{\bibfnamefont{K.}~\bibnamefont{Aihara}},
  \bibinfo{journal}{Chaos} \textbf{\bibinfo{volume}{16}},
  \bibinfo{pages}{013113} (\bibinfo{year}{2006}).

\bibitem[{\citenamefont{Garc\'{\i}a-Ojalvo and Strogatz}(2004)}]{Ojalvo93}
\bibinfo{author}{\bibfnamefont{M.}~\bibnamefont{Garc\'{\i}a-Ojalvo},
  \bibfnamefont{J.~Elowitz}} \bibnamefont{and}
  \bibinfo{author}{\bibfnamefont{H.}~\bibnamefont{Strogatz}},
  \bibinfo{journal}{PNAS} \textbf{\bibinfo{volume}{101}},
  \bibinfo{pages}{10955} (\bibinfo{year}{2004}),
  \urlprefix\url{www.pnas.org/cgi/doi/10.1073/pnas.0307095101}.

\bibitem[{\citenamefont{Cazelles}(1998)}]{Cazelles98}
\bibinfo{author}{\bibfnamefont{B.}~\bibnamefont{Cazelles}},
  \bibinfo{journal}{International Journal of Bifurcations and Chaos}
  \textbf{\bibinfo{volume}{8}}, \bibinfo{pages}{1821 } (\bibinfo{year}{1998}).

\bibitem[{\citenamefont{G\"{u}emez and Mat\'{i}as}(1996)}]{Guemez96}
\bibinfo{author}{\bibfnamefont{J.~J.} \bibnamefont{G\"{u}emez}}
  \bibnamefont{and} \bibinfo{author}{\bibfnamefont{M.~A.}
  \bibnamefont{Mat\'{i}as}}, \bibinfo{journal}{Physica D}
  \textbf{\bibinfo{volume}{96}}, \bibinfo{pages}{334} (\bibinfo{year}{1996}).

\end{thebibliography}

\end{document}